\documentclass[preprint,showpacs,showkeys]{revtex4-1}
\usepackage{graphicx}
\usepackage[caption=false]{subfig}
\usepackage{graphicx,amsmath}
\usepackage{epsfig}
\usepackage{amssymb}
\usepackage{amsmath}
\usepackage{lipsum}
\usepackage{subfig}
\RequirePackage[colorlinks,citecolor=blue,urlcolor=magenta,linkcolor=magenta]{hyperref}
\begin{document} 
\title{Momentum Fractions carried by quarks and gluons in models of proton structure functions at small \textit{x}}
\author{D K Choudhury$^1$}
\author{Baishali Saikia$^1$}
\email[Corresponding author : \ ]{baishalipiks@gmail.com}
\author{K Kalita$^1$}
\affiliation{$^1$Department of Physics, Gauhati University , Guwahati- 781 014, Assam, India}

\begin{abstract}
The paper reports analysis of momentum fractions carried by quarks and gluons in models of Proton structure functions at small \textit{x}. First, we analyze the model proposed by Lastovicka based on self-similarity sometime back. We then make a similar analysis for a second model based on the  same notion which is also free from singularity in \textit{x} : $0<x<1$. The  predictions of both the models are then compared with a recent QCD based Froissart bound compatible model of proton structure function at small \textit{x}, suggested by Block, Durand, Ha and McKay. The results are then compared with the corresponding study in perturbative and Lattice QCD.
\end{abstract}


\maketitle

\section{Introduction}
\label{intro}
\label{A}

How the quarks and gluons share their longitudinal momentum in proton is an important topic of study by itself. It has been studied in \cite{dj, hd, al, rg, ch,ji} within perturbative QCD and in Lattice QCD \cite{md}. It is equally interesting to study the corresponding pattern of such momentum fractions in other phenomenological models of proton \cite{b1,b2,b3,b4,b5,b6}, available in current literature.

One such model is that of Lastovicka \cite{Last} based on self-similarity \cite{bb} at small \textit{x}. While self-similarity is not yet formally established in QCD, it is obtained in renormalization group analysis \cite{kro} and has found its successful phenomenological applications in multi particle hadron physics \cite{5,6,7,8}, since 1980's.

Phenomenological validity range of the model of Ref\cite{Last} is rather limited, $6.2\times10^{-7}\leqslant x\leqslant10^{-2}$ and $0.045\leqslant Q^2\leqslant120$ GeV$^2$. In Ref\cite{DK4}, such pattern was studied assuming the validity in entire \textit{x}-range $0<\textit{x}<1$, while in Ref\cite{DK5}, it was analyzed for \textit{x}$_a<\textit{x}<\textit{x}_b$ ; \textit{x}$_a= 6.2\times 10^{-7}$ and \textit{x}$_b=10^{-2}$ where momentum fraction carried by quarks ($\langle \hat{x}\rangle_q$) and the upper bound of gluons ($\langle \hat{x}\rangle_g$) were obtained for partons having momentum fraction between \textit{x}$_a$ and \textit{x}$_b$. The main reason behind the work of Ref\cite{DK5} is that it is more reasonable to study the model in the phenomenologically allowed range of \textit{x} than going beyond it, as the model\cite{Last} has a singularity at $x\sim 0.019$, outside the range of validity.

One limitation of Ref\cite{DK4,DK5} was that the analytical expression for $\langle \hat{x}\rangle_q$ contains two infinite series ; one in $z \sim \log \frac{1}{x}$ and the other in $\mu_1 \sim \log Q^2$, but only the leading term of each series was considered without studying their convergence properties. This might make the result unstable and unreliable. 

One of the aims of the present communication is to make a re-analysis of Ref\cite{DK5}, and critically examine its stability from the point of view of convergence properties of the infinite series involved. We will then use semi analytical, as well as numerical method and obtain stable values.

In order to remove the undesirable singularity at $x_0\sim 0.019$ in the model of Ref\cite{Last}, an alternative singularity free model \cite{bs}  was suggested but was found to be valid only in a much narrower kinematical range $0.85\leq Q^2\leq 10$ Ge$V^2$. We report the corresponding predictions of small \textit{x} momentum fractions in the model, using the improved numerical method.

The results of both the models are then compared with the corresponding predictions of the QCD based and Froissart bound \cite{fe} compatible phenomenological model suggested by Block, Durand, Ha and McKay \cite{blo}, which has an explicit $x$ and $Q^2$- dependent structure function. The results are then compared with the perturbative, as well as Lattice QCD.

In section \ref{B}, we outline the essential formalism and the model of Ref\cite{Last} : Model 1 and also Model 2 \cite{bs}. In section \ref{C}, we report the essential features of the model of Ref\cite{blo} : Model 3. Section \ref{D} contains the conclusions.

\section{Formalism}
\label{B}
\subsection{Proton structure function based on self-similarity}
The self-similarity based model of the proton structure function of Ref\cite{Last} is based on \textit{x} and $Q^2$ parton distribution function(PDF) $q_i(x,Q^2)$. Choosing the magnification factors $M_1= \left(1+\dfrac{Q^2}{Q_0^2}\right)$ and $M_2= \left(\dfrac{1}{x}\right)$, the unintegrated parton density function (uPDF) can be written as \cite{Last}

\begin{equation}
\label{E1}
\log[M^2.f_i(x,Q^2)]= D_1.\log\frac{1}{x}.\log\left(1+\frac{Q^2}{Q_0^2}\right)+D_2.\log\frac{1}{x}+D_3.\log\left(1+\frac{Q^2}{Q_0^2}\right)+D_0^i
\end{equation}
\\
where \textit{i} denotes a quark flavor. Here $D_1,\ D_2,\ D_3$ are the three flavor independent model parameters while $D_0^i$ is the only flavor dependent normalization constant. $M^2$(=1 GeV$^2$) is introduced to make (PDF) $q_i(x,Q^2)$, as defined below (in Eqn \ref{E2}), dimensionless. The integrated quark densities then can be defined as

\begin{equation}
\label{E2}
q_i(x,Q^2) = \int_0^{Q^2}f_i(x,Q^2)dQ^2
\end{equation}
\\
As a result, the following analytical parametrization of a quark density is obtained by using Eqn(\ref{E2}) \cite{DK5} : \textbf{(Model 1)}

\begin{equation}
\label{E3}
q_i(x,Q^2) = e^{D_0^i}f(x,Q^2)
\end{equation}
where
\begin{equation}
\label{E4}
f(x,Q^2)= \frac{Q_0^2 \ \left( \dfrac{1}{x}\right) ^{D_2}}{M^2\left(1+D_3+D_1\log\left(\dfrac{1}{x}\right)\right)} \left(\left(\frac{1}{x}\right)^{D_1\log \left(1+\frac{Q^2}{Q_0^2}\right)} \left(1+\frac{Q^2}{Q_0^2}\right)^{D_3+1}-1 \right)
\end{equation}
\\
is flavor independent. Using Eqn(\ref{E3}) in the usual definition of the structure function $F_2(x,Q^2)$, one can get
\begin{equation}
\label{E5}
F_2(x,Q^2)=x\sum_i e_i^2 \left( q_i(x,Q^2)+ \bar{q}_i(x,Q^2)\right) 
\end{equation}
or it can be written as
\begin{equation}
\label{E6}
F_2(x,Q^2)=e^{D_0}xf(x,Q^2)
\end{equation}
\\
where 
\begin{equation}
\label{Ea}
e^{{D_0}}=\sum_{i=1}^{n_f}e^{2}_{i}\left(e^{D_0^i}+ e^{\bar{D}_0^i}\right)
\end{equation}
\\
Eqn(\ref{E5}) involves both quarks and anti-quarks. As in Ref\cite{Last}, we assume the same parametrization both for quarks and anti-quarks. Assuming the quark and anti-quark have equal normalization constants, we obtain for a specific flavor
\begin{equation}
\label{Eb}
e^{{D_0}}=\sum_{i=1}^{n_f}e^{2}_{i}\left(2 e^{D_0^i}\right)
\end{equation}
\\
It shows that the value of $D_0$ will increase as more and more number of flavors contribute to the structure function.\\ With $n_f=3 , 4 $ and 5, it reads explicitly as
\begin{eqnarray}
\label{Ec}
n_f=3 &:& \ \ e^{D_0}= 2 \left( \frac{4}{9}e^{{D_0}^{u}}+\frac{1}{9}e^{{D_0}^{d}}+\frac{1}{9}e^{{D_0}^{s}} \right) \\
\label{Ed}
n_f=4 &:& \ \ e^{D_0}= 2 \left(\frac{4}{9}e^{{D_0}^{u}}+\frac{1}{9}e^{{D_0}^{d}}+\frac{1}{9}e^{{D_0}^{s}}+\frac{4}{9}e^{{D_0}^{c}}\right) \\
\label{Ef}
n_f=5 &:& \ \ e^{D_0}= 2 \left(\frac{4}{9}e^{{D_0}^{u}}+\frac{1}{9}e^{{D_0}^{d}}+\frac{1}{9}e^{{D_0}^{s}}+\frac{4}{9}e^{{D_0}^{c}}+\frac{1}{9}e^{{D_0}^{b}}\right) 
\end{eqnarray}
\\
Since each term of right hand sides of Eqn(\ref{Ec}),(\ref{Ed}), and (\ref{Ef}) is positive definite, it is clear, the measured value of $D_0$ increases as $n_f$ increases. However, single experimentally determined parameter $D_0$ can not ascertain the individual contribution from various flavors.
\\ \\
From HERA data \cite{H1,ZE}, Eqn(\ref{E6}) was fitted in Ref\cite{Last} with
\begin{eqnarray}
\label{E7}
D_0 &=& 0.339\pm 0.145 \nonumber \\
D_1 &=& 0.073\pm 0.001 \nonumber \\
D_2 &=& 1.013\pm 0.01 \nonumber \\
D_3 &=& -1.287\pm 0.01 \nonumber \\
Q_0^2 &=& 0.062\pm 0.01 \ {\text G\text e\text V^2}
\end{eqnarray}

in the kinematical region,
\begin{eqnarray}
\label{E8}
& & 6.2\times10^{-7}\leq x\leq 10^{-2} \nonumber \\
& & 0.045\leq Q^2 \leq 120 \ {\text G\text e\text V^2}
\end{eqnarray}

\subsection{Singularity free structure function: Model 2}
The defining equations of the model of Ref\cite{Last} (Eqn \ref{E1}-\ref{E4} above) do not ascertain the numerical values and signs of the parameters $D_j$ s. These are determined from data\cite{H1,ZE}, leading to the set of Eqn(\ref{E7}) in the kinematic range (Eqn \ref{E8}). However, the phenomenological analysis has one inherent limitation: Due to the  negative value of $D_3$, Eqn(\ref{E6}) develops a singularity at $x_0 \backsim 0.019$ \cite{DK4, DK5}, as it satisfies the condition $1+D_3+D_1\log\dfrac{1}{x_0}=0$, contrary to the expectation of a physically viable form of structure function. We, therefore, explore the possibility of an alternate model which is singularity free.

Redefining the model parameters $D_j$ s by $D'_j$ s (\textit{j}=1,2,3) and (PDF) $q_i(x,Q^2)$ by $q'_i(x,Q^2)$ and also structure function $F_2(x,Q^2)$ by $F'_2(x,Q^2)$ in the present model, we get the following forms of PDF and structure function as : \textbf{(Model 2)}

\begin{equation}
\label{a1}
q_i'(x,Q^2)=\frac{e^{D_0'^i}\ Q_0'^2\ \left( \dfrac{1}{x}\right) ^{D'_2}}{M^2 \left(1+D'_3+D'_1\log\dfrac{1}{x}\right)} \left(\left(\frac{1}{x}\right)^{D'_1\log\left(1+\frac{Q^2}{Q_0'^2}\right)}\left(1+\frac{Q^2}{Q_0'^2}\right)^{D'_3+1}-1\right)
\end{equation}
and
\begin{equation}
\label{a2}
F'_2(x,Q^2)=\frac{e^{D'_0}\ Q_0'^2\ \left( \dfrac{1}{x}\right) ^{D'_2-1}}{M^2 \left(1+D'_3+D'_1\log\dfrac{1}{x}\right)} \left(\left(\frac{1}{x}\right)^{D'_1\log\left(1+\frac{Q^2}{Q_0'^2}\right)}\left(1+\frac{Q^2}{Q_0'^2}\right)^{D'_3+1}-1\right)
\end{equation}
respectively.

\begin{table}[!tbp]
\caption{\label{Table4}%
Results of the fit of \textbf{Model 2}, Eqn(15)}
\begin{ruledtabular}
\begin{tabular}{ccccccc}
\textrm{$D'_0$}&
\textrm{$D'_1$}&
\textrm{$D'_2$}&
\textrm{$D'_3$}&
\textrm{$Q_0'^2$(GeV$^2$)}&
\textrm{$\chi^2$}&
\textrm{$\chi^2$/ndf} \\
\colrule
-2.971\tiny${\pm 0.409}$ & 0.065\tiny${\pm 0.0003}$ & 1.021\tiny${\pm 0.004}$ & 0.0003\tiny${\pm 0.0001}$ & 0.20\tiny${\pm 0.0008}$ & 18.829 & 0.20\\
\end{tabular}
\end{ruledtabular}
\end{table}

The model parameters $\left(D'_0 , D'_1 , D'_2 , D'_3 , Q_0'^2 \right)$ are determined \cite{bs} by using the compiled HERA data \cite{HERA}, instead of earlier data \cite{H1,ZE}, used in Ref\cite{Last} and obtained more restrictive range of $Q^2$ and \textit{x} : $0.85\leq Q^2 \leq$ 10 GeV$^2$ and $2\times10^{-5}\leq x\leq 0.02$ respectively with the  fitted parameters given in the Table \ref{Table4}.

\begin{figure*}[!tbp]
 \captionsetup[subfigure]{labelformat=empty}
 \centering
  \subfloat[]{\includegraphics[width=.3\textwidth]{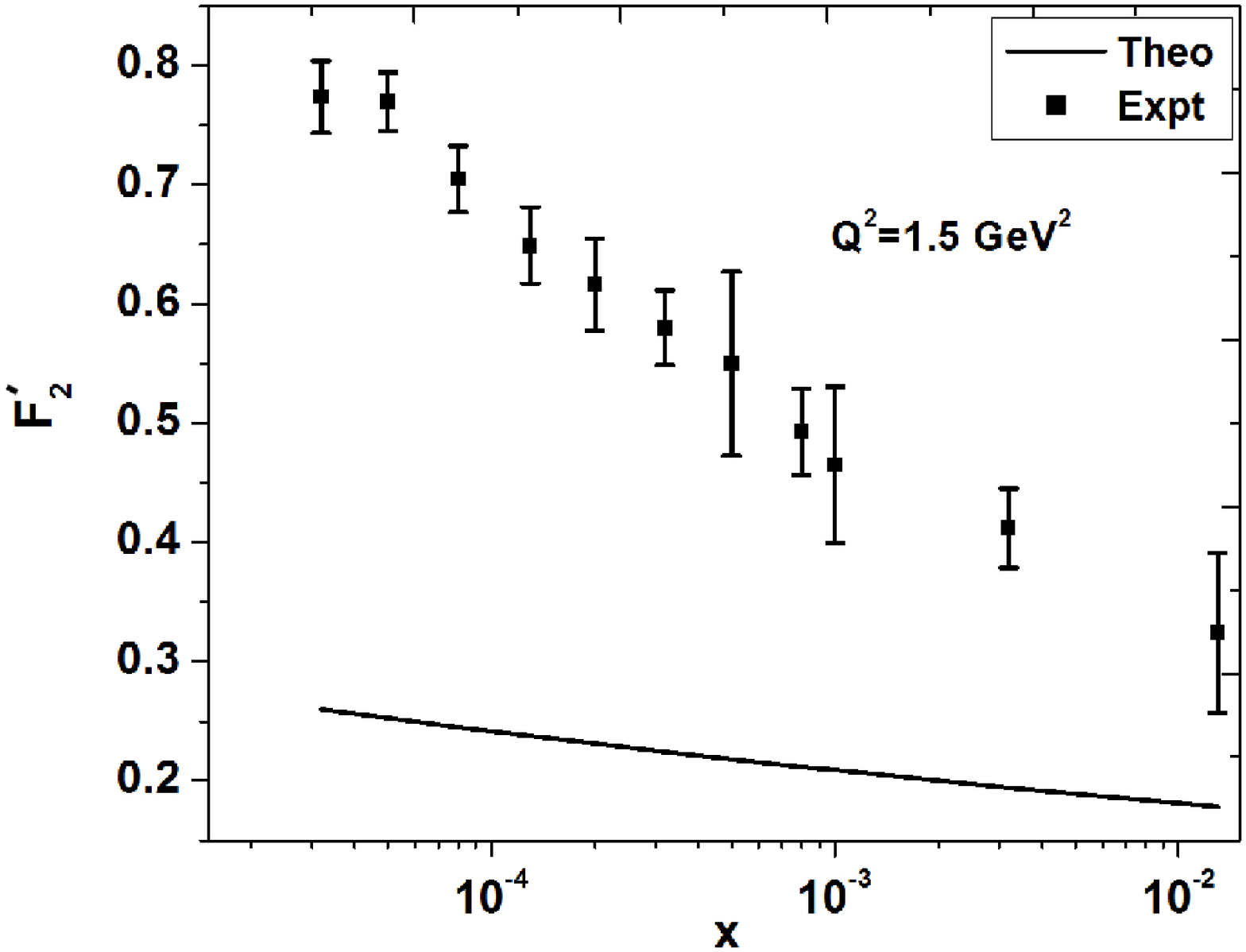}}\quad
\subfloat[]{\includegraphics[width=.3\textwidth]{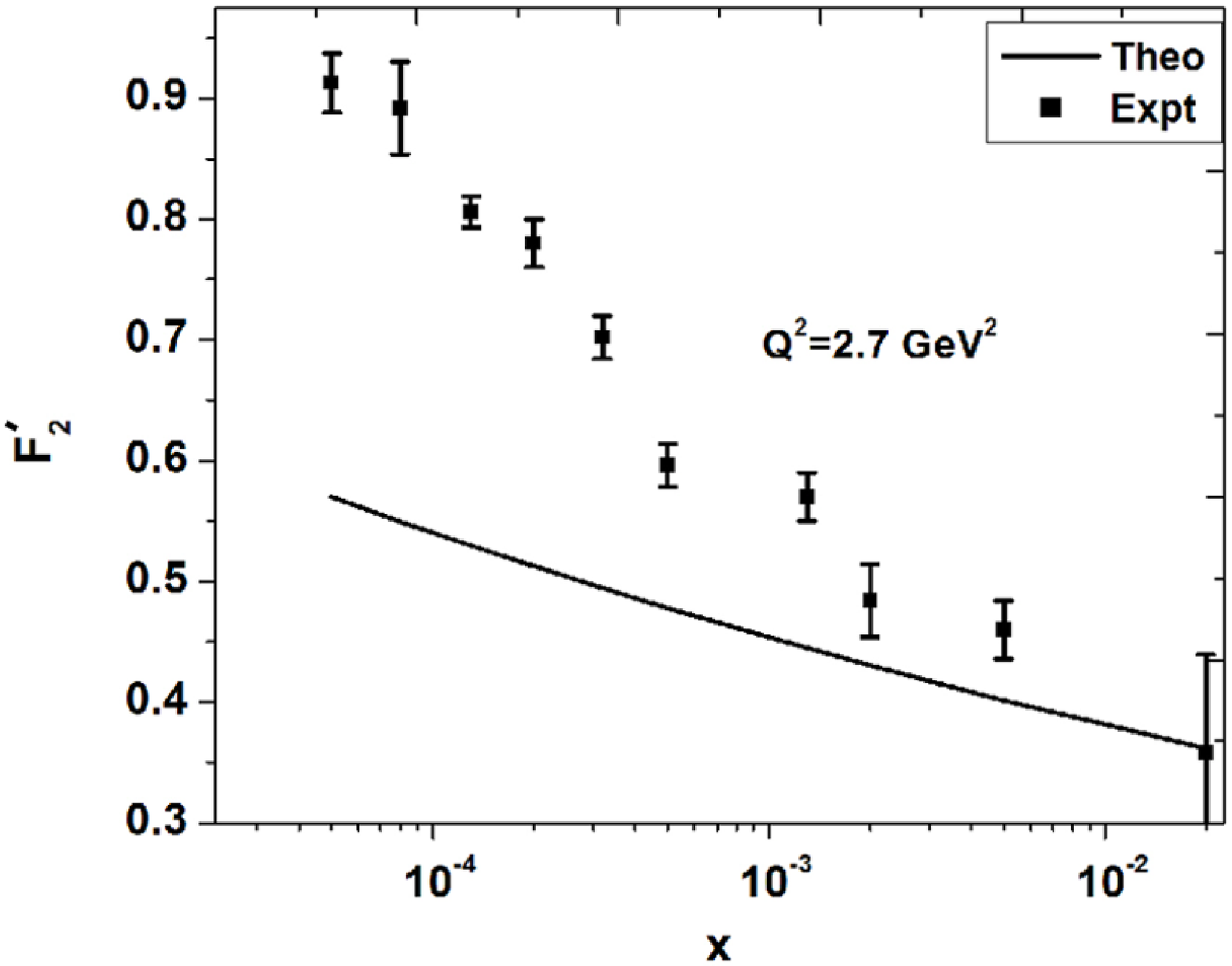}}\quad
 \subfloat[]{\includegraphics[width=.3\textwidth]{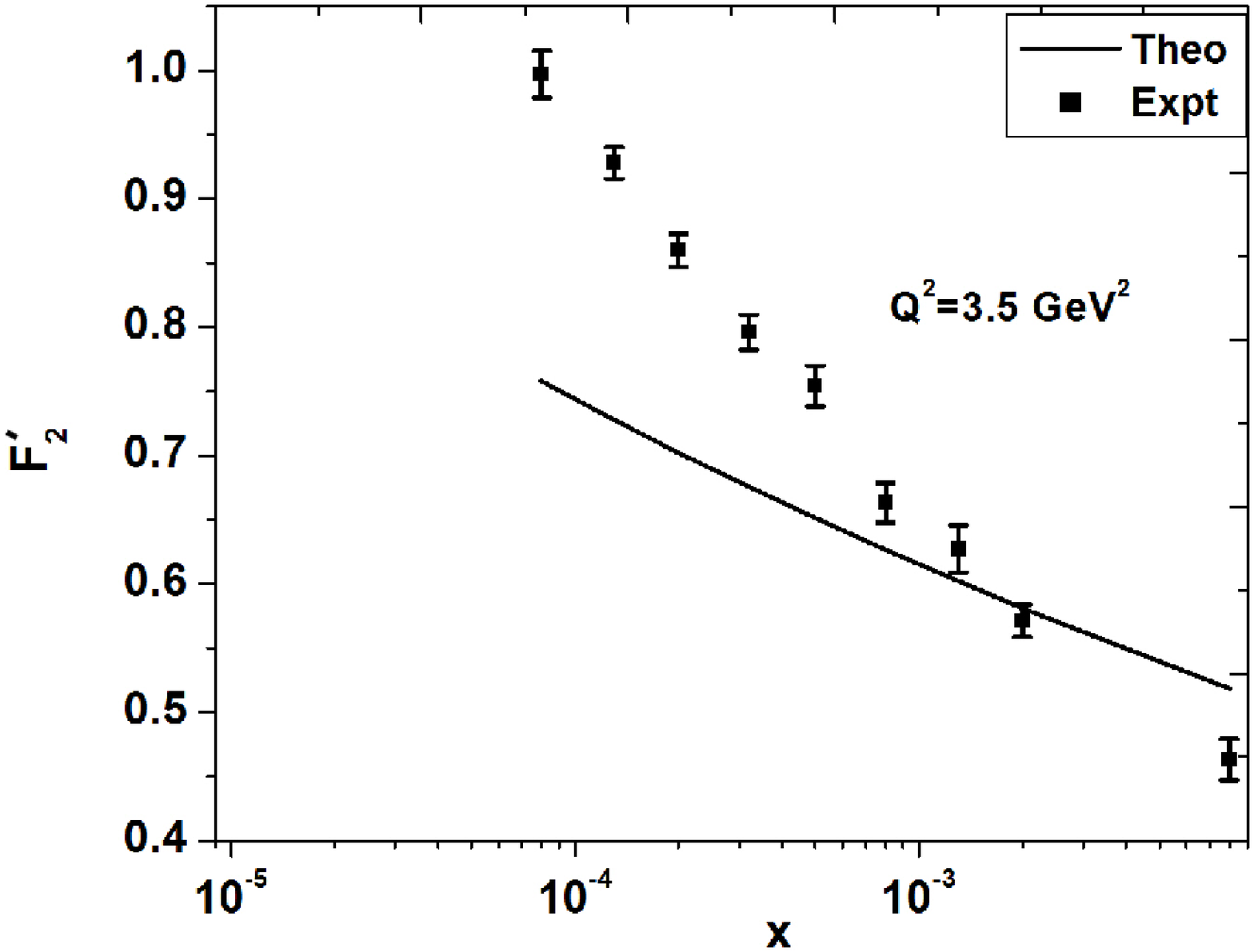}}\quad
\subfloat[]{\includegraphics[width=.3\textwidth]{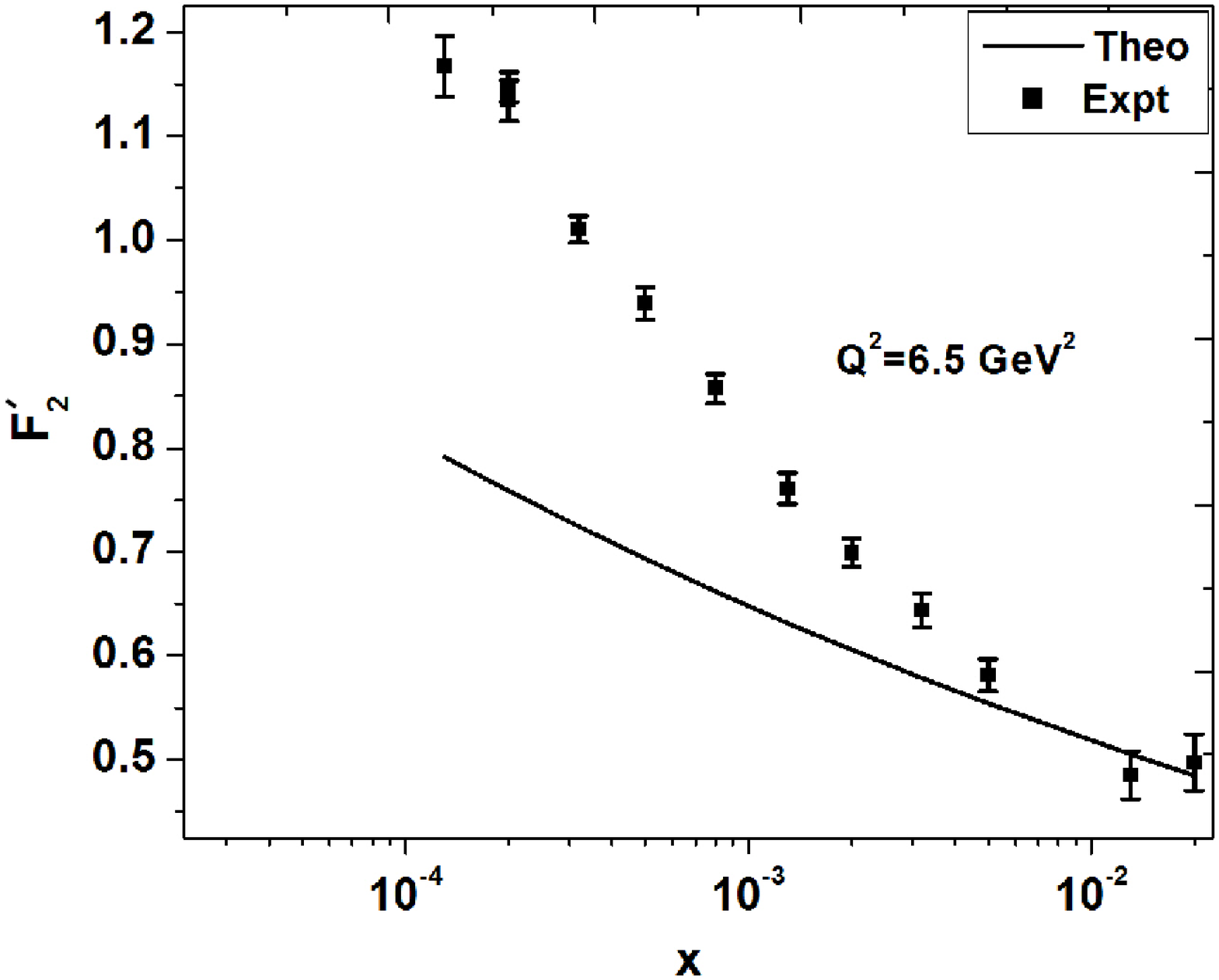}}\quad
\subfloat[]{\includegraphics[width=.3\textwidth]{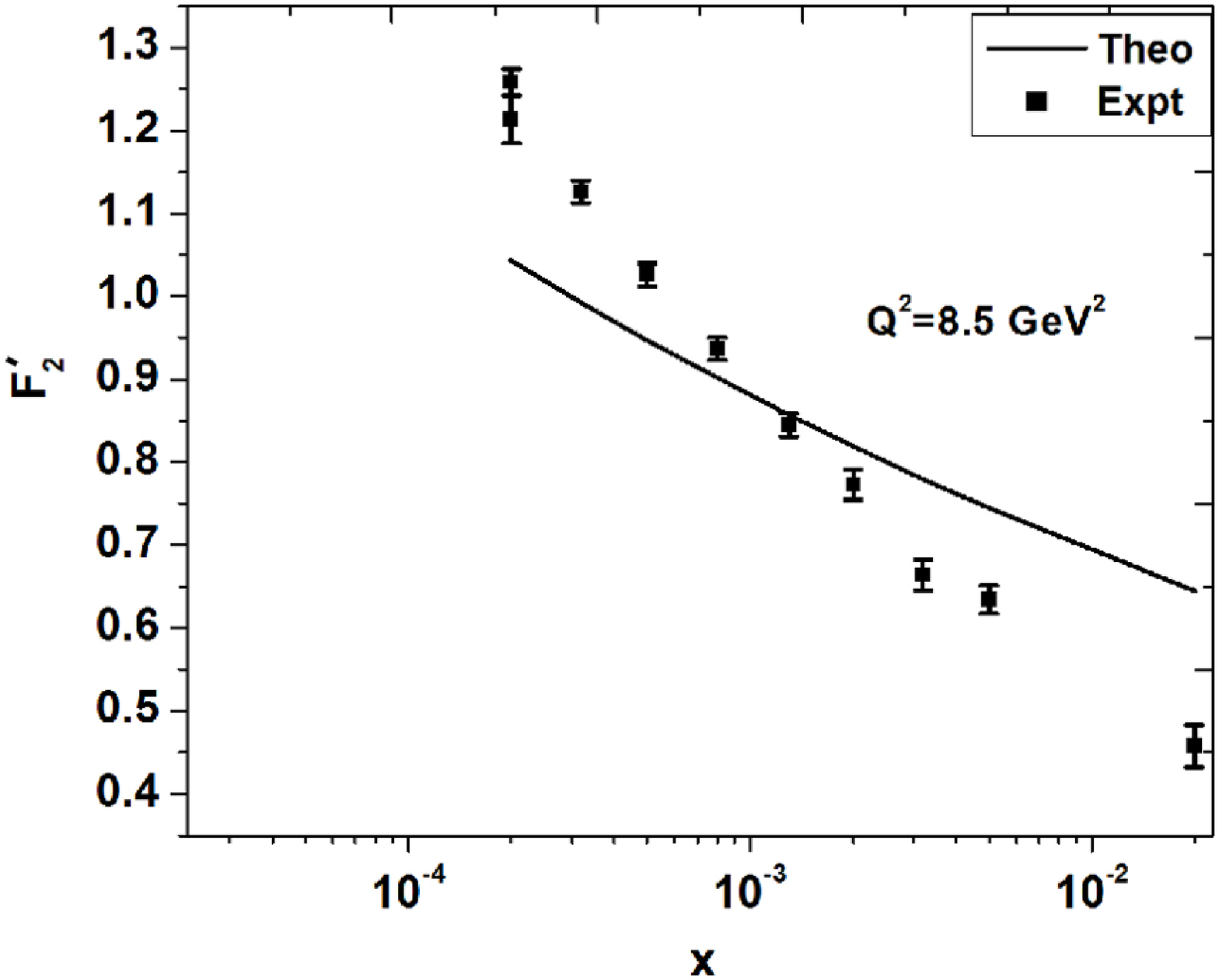}}\quad
 \subfloat[]{\includegraphics[width=.3\textwidth]{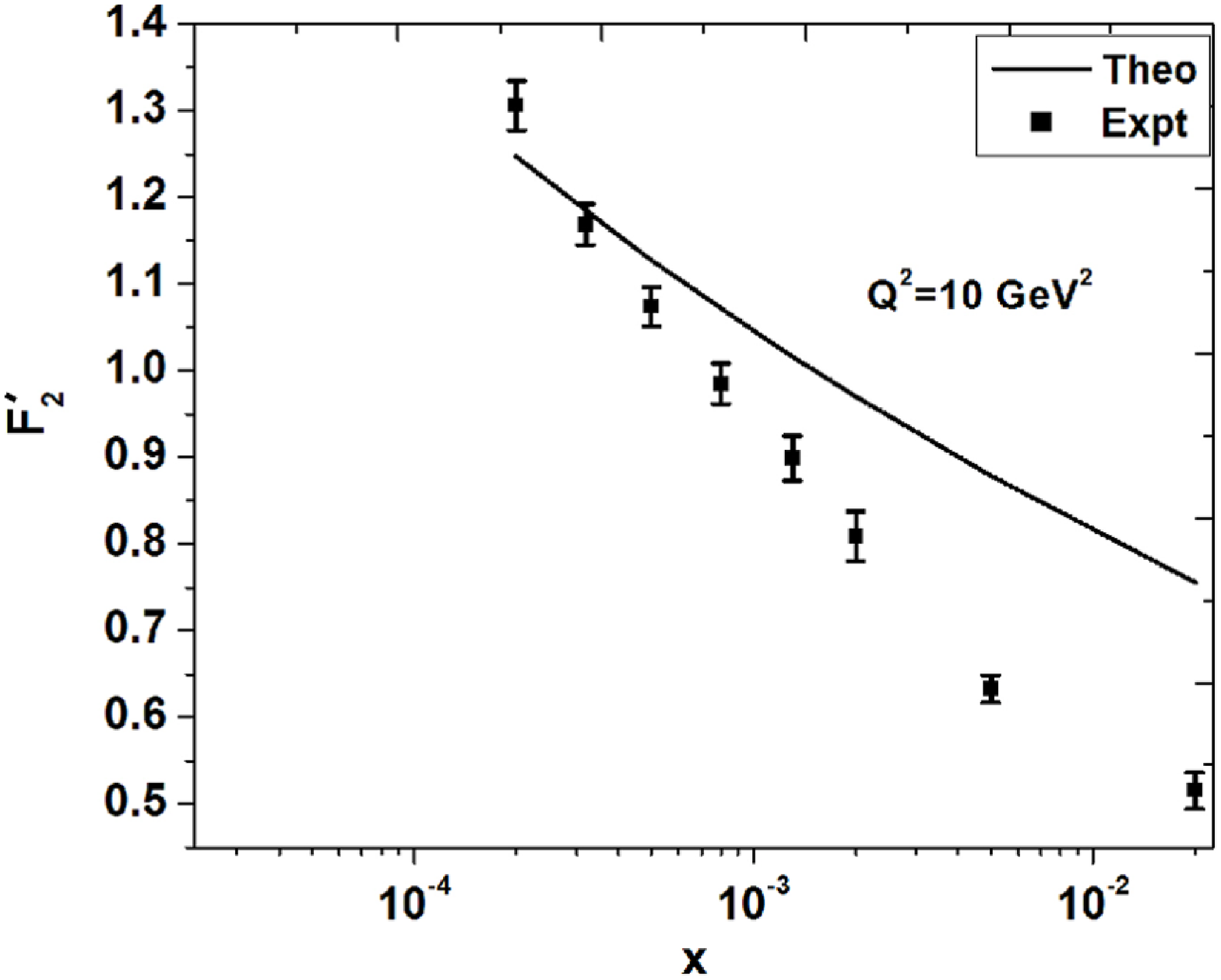}}\quad
\caption{{\footnotesize comparison of the structure function $F'_2$ of \textbf{Model 2} as a function of $x$ in bins of $Q^2$ with measured data of $F_2$  from HERAPDF1.0\cite{HERA}}}
 \label{Fig1}
\end{figure*}

In Fig \ref{Fig1}, we plot $F'_2$ of Model 2 as a function of $x$ for six representative values of $Q^2$ ($Q^2$= 1.5, 2.7, 3.5, 6.5, 8.5, 10 GeV$^2$) within the phenomenologically allowed range; 0.85 $\leq Q^2 \leq$ 10 GeV$^2$. We also show the corresponding available data from Ref\cite{HERA}.

It shows that as the model parameters have constraint for the positivity alone, the range of validity shrinks from $Q^2=120$ GeV$^2$ to $Q^2=10$ GeV$^2$. Thus our analysis indicates that the phenomenological range of validity of the present version of the model is more restrictive; $0.85\leq Q^2 \leq 10$ GeV$^2$ and $2\times 10^{-5} \leq x \leq 0.02$, to be compared with Eqn(\ref{E8}) of the previous version of Ref\cite{Last}. Also, the individual $\chi^2$ at $Q^2= 8.5$ and 10 GeV$^2$ is minimum to be compared with $Q^2= 4.5$ and 10 GeV$^2$, which is quite larger than that of 10 GeV$^2$. It is same for $Q^2= 1.5$ GeV$^2$ too. Basically, our results valid in small area in between $Q^2$ of 8.5 and 10 GeV$^2$, but due to the unavailability of the experimental data points, the difference cant be shown explicitly.

We also observe the following features of the model compared to data: at $Q^2=1.5$ GeV$^2$ data overshoots the theory. But as $Q^2$ increases, the theoretical curve comes closer to data. At $Q^2$=10 GeV$^2$, on the other hand, the theory exceeds data. Main reason of this feature is that the $x$-slope of the model is less than that of the data. Specifically, due to positive $D_3$, the growth of the structure function with $Q^2$ becomes faster as can be seen from Eqn(\ref{E4}) i.e. $$\left( 1+\frac{Q^2}{Q_0'^2}\right) ^{(1+D'_3)} \approx \left( 1+\frac{Q^2}{Q_0'^2}\right) ^{1.0003}$$ at higher values of $Q^2 >$ 1 GeV$^2$ to be compared with $$\left( 1+\frac{Q^2}{Q_0^2}\right) ^{(1+D_3)} \approx \left( 1+\frac{Q^2}{Q_0^2}\right) ^{-0.287}$$ of Ref\cite{Last}.

\subsection{Graphical representation of PDF}
\begin{figure}[!bp]
\subfloat[]{%
  \includegraphics[width=0.48\linewidth]{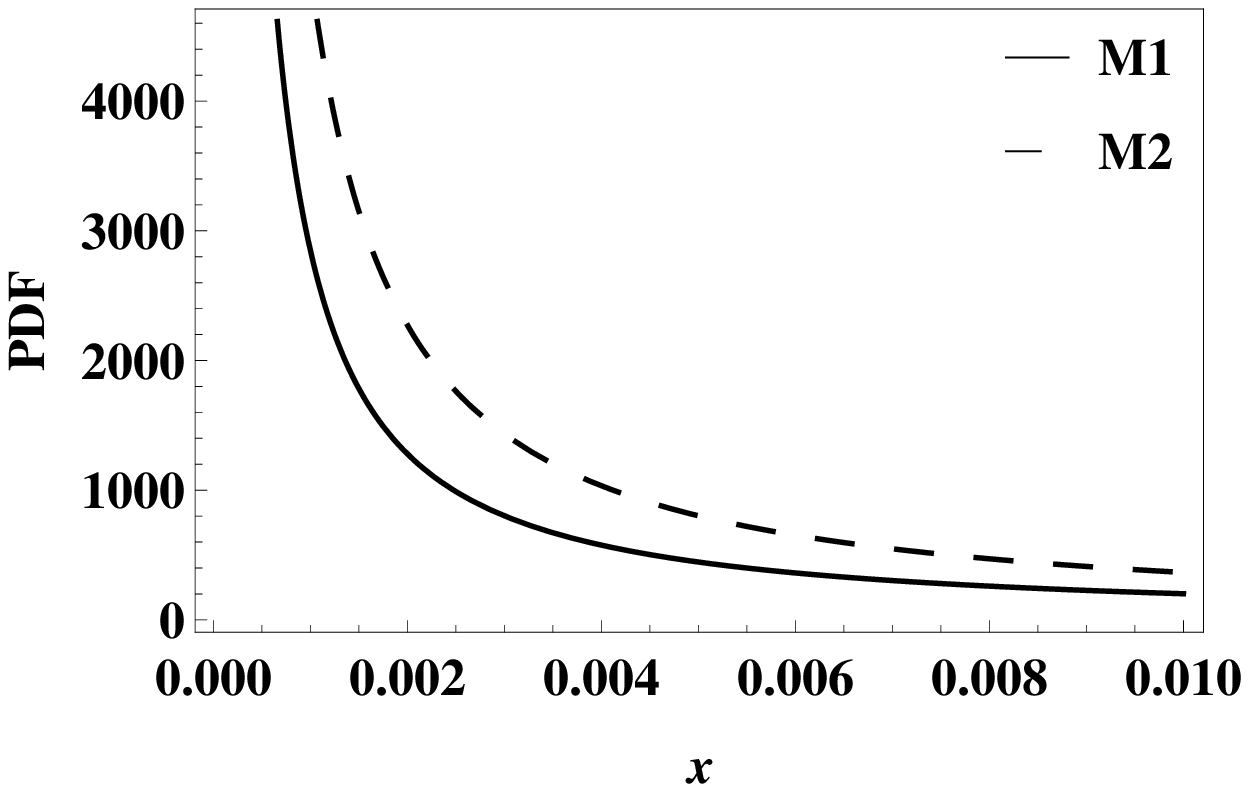}%
}\vspace{1ex}
\subfloat[]{%
  \includegraphics[width=0.50\linewidth]{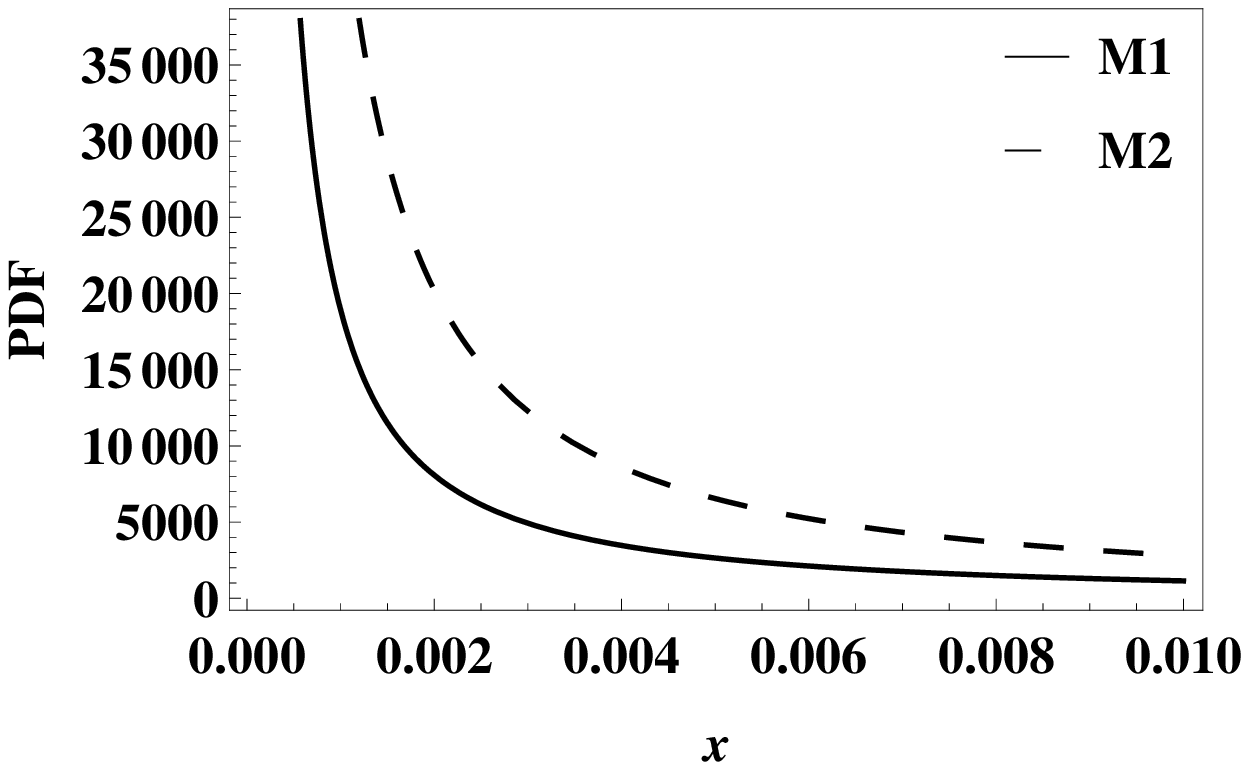}%
}\vspace{1ex}
\caption{PDF vs $x$ for two representative values of (a) $Q^2= 2$ GeV$^2$ and (b) $Q^2= 10$ GeV$^2$ for Models 1 and 2 respectively. Here, M1 (black line) represents the PDF for \textbf{Model 1}. Similarly, M2 (black dashed) represents the PDF for \textbf{Model 2}.}
\label{F2}
\end{figure}

\begin{figure}[!tbp]
\subfloat[]{%
  \includegraphics[width=0.49\linewidth]{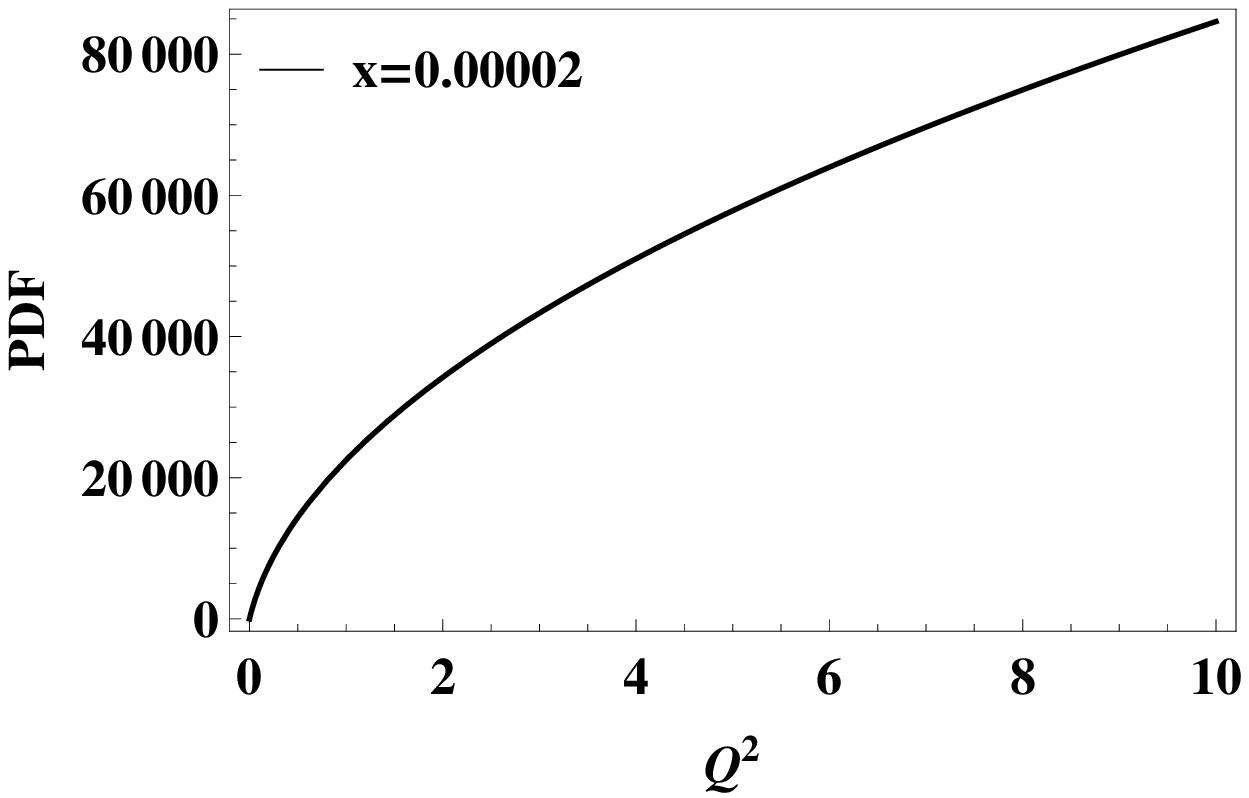}%
}\vspace{1ex}
\subfloat[]{%
  \includegraphics[width=0.47\linewidth]{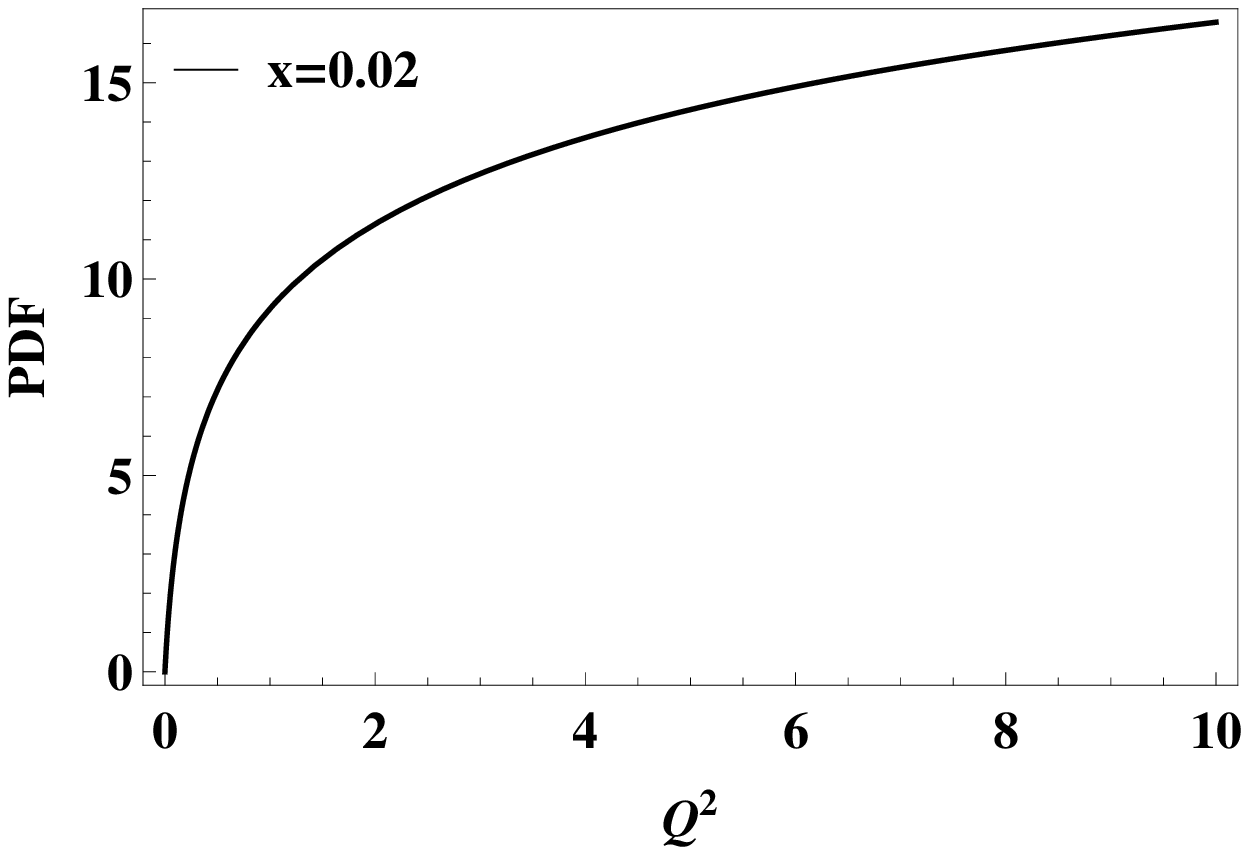}%
}\vspace{1ex}
\caption{PDF vs $Q^2$ for two representative values of (a) $x=2\times10^{-4}$ and (b) $x=0.02$ for\textbf{ Model 1}.}
\label{F3}
\end{figure}

\begin{figure}[!tbp]
\subfloat[]{%
  \includegraphics[width=0.49\linewidth]{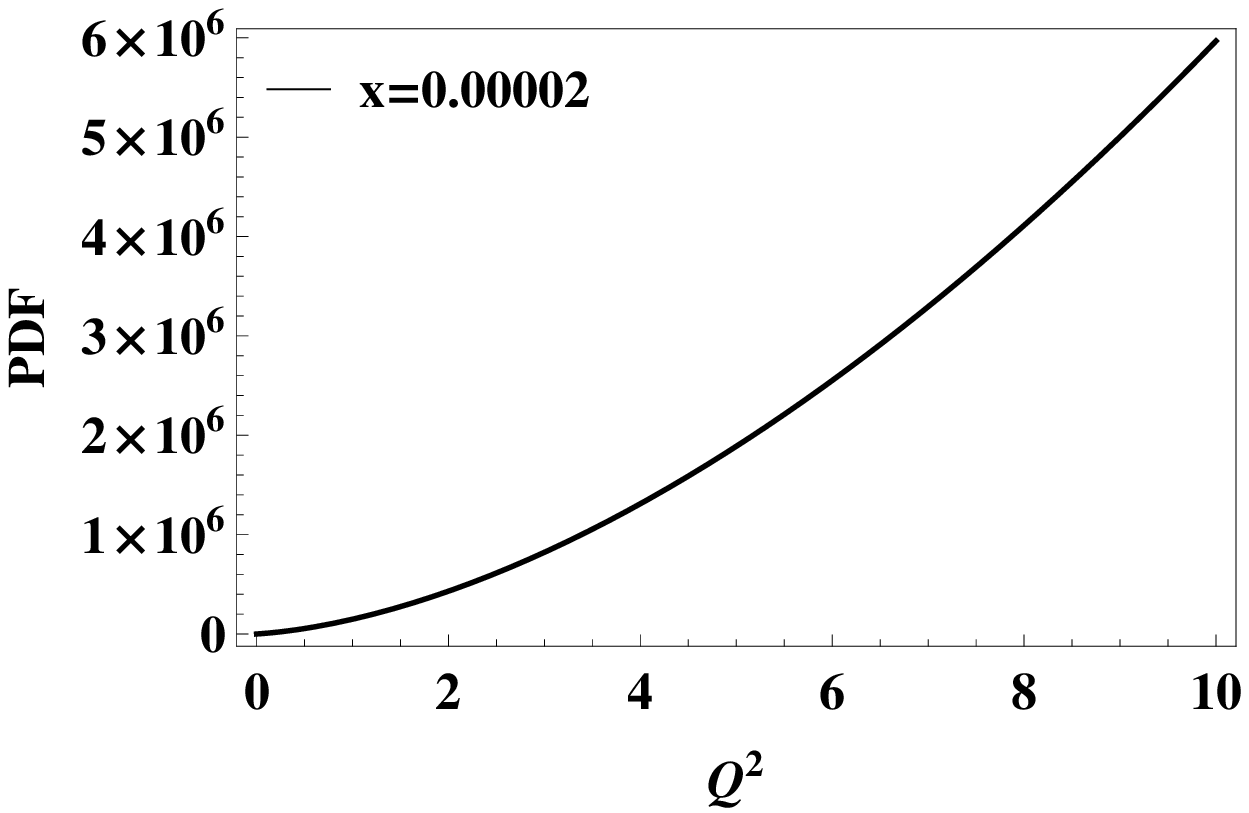}%
}\vspace{1ex}
\subfloat[]{%
  \includegraphics[width=0.47\linewidth]{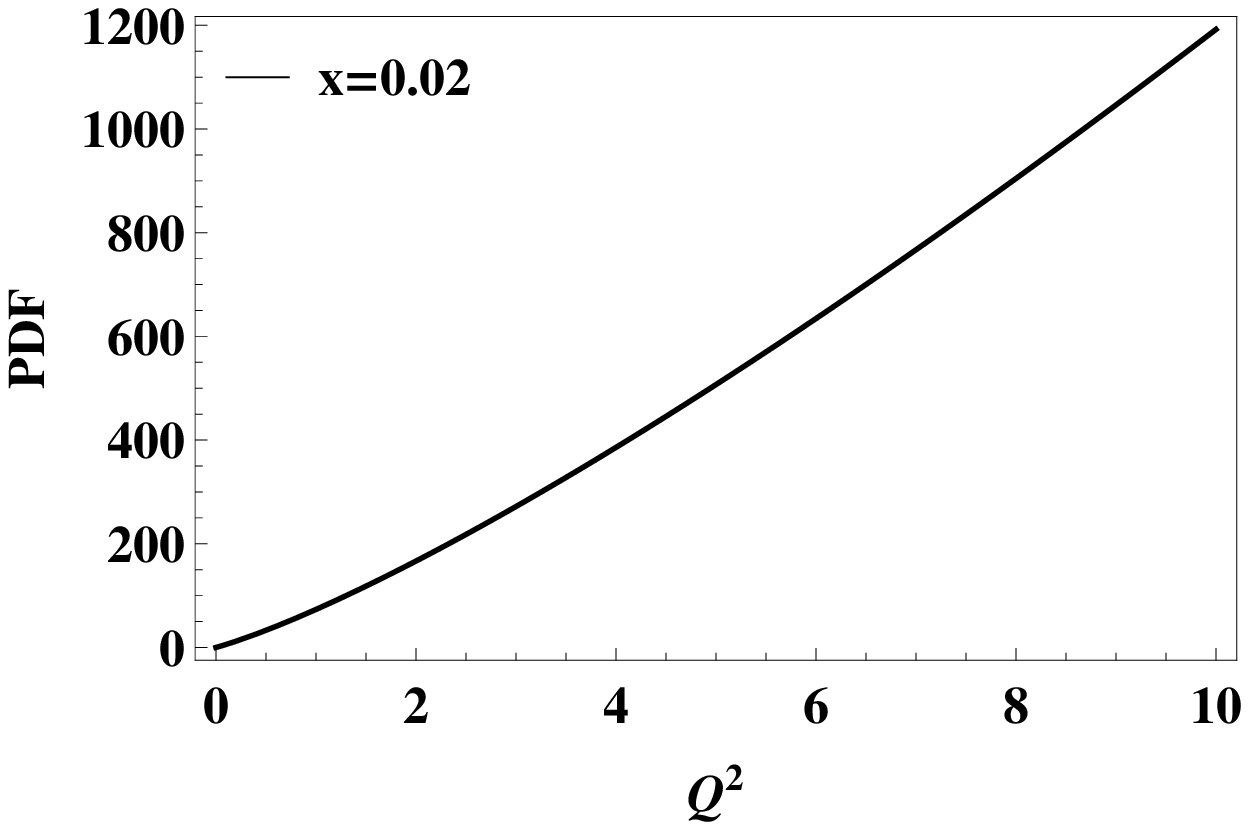}%
}\vspace{1ex}
\caption{PDF vs $Q^2$ for two representative values of (a) $x=2\times10^{-4}$ and (b) $x=0.02$ for \textbf{Model 2}.}
\label{F4}
\end{figure}

From Eqn(\ref{E3}), the form of PDFs for Models 1 and 2 can be written as follows, excluding the flavor dependent term $e^{{D_0}^i}$. 

\begin{equation}
\label{E9}
{\text M\text o\text d\text e\text l \ \text 1: } \quad f(x,Q^2)= \frac{Q_0^2 \ \left( \dfrac{1}{x}\right) ^{D_2}}{M^2\left(1+D_3+D_1\log\left(\dfrac{1}{x}\right)\right)} \left(\left(\frac{1}{x}\right)^{D_1\log \left(1+\frac{Q^2}{Q_0^2}\right)} \left(1+\frac{Q^2}{Q_0^2}\right)^{D_3+1}-1 \right)
\end{equation}
\begin{equation}
\label{E10}
{\text M\text o\text d\text e\text l \ \text 2: } \quad {\footnotesize f(x,Q^2)= \frac{Q_0'^2 \ \left( \dfrac{1}{x}\right) ^{D'_2}}{M^2\left(1+D'_3+D'_1\log\left(\dfrac{1}{x}\right)\right)} \left(\left(\frac{1}{x}\right)^{D'_1\log \left(1+\frac{Q^2}{Q_0'^2}\right)} \left(1+\frac{Q^2}{Q_0'^2}\right)^{D'_3+1}-1 \right)}
\end{equation}
\\
Graphical representation of PDFs of Model 1 and 2 are shown in Fig \ref{F2}, \ref{F3} and \ref{F4}

\subsection{Momentum Sum Rule}
The momentum sum rule is given as\cite{DK4,DK5,clo}
\begin{equation}
\label{E11}
\int_0^1 x\sum \left(q_i(x,Q^2)+\bar{q}_i(x,Q^2)\right)dx+\int_0^1 G(x,Q^2)\ dx=1
\end{equation}
where \begin{equation}
\label{E12}
G(x,Q^2)=xg(x,Q^2)
\end{equation}
$g(x,Q^2)$ is the gluon number density. It can be converted \cite{DK5} into an inequality if the information about quarks and gluons is available only in a limited range of $x$, say $x_a\leq x\leq x_b$ i.e.
\begin{equation}
\label{E13}
\int_{x_a}^{x_b} x\sum \left(q_i(x,Q^2)+\bar{q}_i(x,Q^2)\right)dx+ \int_{x_a}^{x_b} G(x,Q^2)\ dx\ < 1
\end{equation}
We have omitted the equality sign in Eqn(\ref{E13}) because it will correspond to a nucleon, populated by small quarks and gluons (parton) only within the range $x_a<x<x_b$, which makes no sense physically. This yields the respective information when the momentum fractions carried by small $x$ quarks and gluons in $x_a<x<x_b$ to be
\begin{equation}
\label{E14}
\langle \hat{x}\rangle_q =\int_{x_a}^{x_b}x\sum \left(q_i(x,Q^2)+\bar{q}_i(x,Q^2)\right)dx
\end{equation}
Using Eqn(\ref{E5}), we can write
\begin{equation}
\label{E15}
\langle\hat{x}\rangle_q = \left( \sum_{i=1}^{n_f}e^{2}_{i} \right)^{-1} \int_{x_a}^{x_b} F_{2}(x,Q^2)dx
\end{equation}

and \begin{equation}
\label{E16}
\langle \hat{x}\rangle_g < \int_{x_a}^{x_b} G(x,Q^2)\ dx \ < 1-\langle \hat{x}\rangle_q
\end{equation}
Note that Eqn(\ref{E5}) yields only the upper limit of the fractional momentum carried by the gluons in the regime $x_a<x<x_b$.\\
\\
In terms of structure function, the momentum sum rule inequality is 
\begin{equation}
\label{E17}
\int_{x_a}^{x_b}\left\lbrace a F_{2}(x,Q^2)+ G(x,Q^2)\right\rbrace dx\ < 1
\end{equation}
where $a=\frac{e^{\tilde{D_0}}}{e^{D_0}}$ is $Q^2$-independent parameter, determined from data \cite{a}, $a=3.1418$ \cite{DK4}, using the fractionally charged quarks.

\subsection{Analytical Expression of $\langle \hat{x}\rangle_q$ of Model 1 and its limitations:}
The analytical expression of $\langle \hat{x}\rangle_q$ is given as (Eqn 23 of Ref\cite{DK5})
\begin{equation}
\label{E18}
\langle \hat{x}\rangle_q=\frac{e^{\tilde{D_0}} Q_0^2}{D_{1} M^2} e^{\left(\frac{1+D_3}{D_1}\right)(2-D_2)}\left\lbrace \left( 1+\frac{Q^2}{Q_0^2}\right)^{D_3+1} e^{-\left(\frac{1+D_3}{D_1}\right)D_1\log \left(1+\frac{Q^2}{Q_0^2}\right)}I_1-I_2\right\rbrace 
\end{equation}
Where the integrals $I_1$ and $I_2$ are expressible in terms of infinite series
\begin{equation}
\label{E19}
I_i=\int\frac{e^{\mu_{i}z}}{z}dz=\log|z|+\sum_{n=1}^{\infty}\frac{\mu_{i}^n z^n}{n.n!} \ \ \, \ \ \ , \,\,\, i=1,2
\end{equation}
where
\begin{equation}
\label{E20}
z=\frac{1+D_3}{D_1}+\log\frac{1}{x}
\end{equation}
and
\begin{equation}
\label{E21}
\mu_1 = D_1 \log \left(1+\frac{Q^2}{Q_0^2}\right)+D_2 -1
\end{equation}
\begin{equation}
\label{E22}
\mu_2 = D_2-1
\end{equation}

In Ref\cite{DK5}, only the 1st term of the infinite series is taken into account without taking into account the convergence property and their $Q^2$-dependence. Below, we address to this point.

\section{Results and Discussion}
\label{C}
\subsection{$Q^2$-dependence of the convergence of the infinite series of Model 1:}
The integral $I_1$ is $Q^2$-dependent while $I_2$ is not, as can be seen from Eqn (\ref{E21}) and (\ref{E22}) above respectively. Convergent condition between n$^{th}$ and (n-1)$^{th}$ term of the infinite series is 
\begin{equation}
\label{E23}
\frac{\mu_{i}^{(n-1)}.\,z^{(n-1)}}{(n-1).(n-1)!} \gg \frac{\mu_{i}^n.\,z^n}{n.n!} \ \ \ ;\ \ i=1,2
\end{equation}
leading to
\begin{equation}
\label{E24}
z \ll \frac{n^2}{(n-1)}.\frac{1}{\mu}
\end{equation}

It can be explicitly seen that if one includes more and more terms in the infinite series $I_1$, the convergent condition shifts to higher values of $Q^2$. As an illustration, the relative convergence taking respectively the ratios of the 3rd vs 2nd term, 4th vs 3rd term, 5th vs 4th, 6th vs 5th term results in the inequalities as 
$$\log \left(1+\frac{Q^2}{Q_0^2}\right) \ll 15.384 \ \ \ \ \ \ \ \ \ \ \ \ \ \ 19(a)$$
$$\ \ \ \ \ \ \ \ \ \ \ \ \ \ \ \ \ \ \ \ll 45.454 \ \ \ \ \ \ \ \ \ \ \ \ \ \ 19(b)$$
$$\ \ \ \ \ \ \ \ \ \ \ \ \ \ \ \ \ \ \ \ll 52.631 \ \ \ \ \ \ \ \ \ \ \ \ \ \ 19(c)$$
$$\ \ \ \ \ \ \ \ \ \ \ \ \ \ \ \ \ \ \ \ll 58.823 \ \ \ \ \ \ \ \ \ \ \ \ \ \ 19(d)$$

In Model 1, these inequalities saturates at 2.9$\times10^5$, 3.4$\times10^{18}$, 4.4$\times10^{21}$, 2.1$\times10^{24}$ GeV$^2$ respectively, which are far above the phenomenological range of validity in $Q^2$ $\leqslant$ 120 GeV$^2$, as well as the experimentally accessible HERA range 3$\times10^4$ GeV$^2$ \cite{HERA}. However, it is the slow convergence of the two infinite series, which might make the result highly unstable.

\begin{table}[!tbp]
\caption{}\label{t1}%
Values of $\langle\hat{x}\rangle_q$ of \textbf{Model 1} with higher order terms in $I_1$ and $I_2$ for different $Q^2$
\begin{center}
\begin{tabular}{|c|c|c|c|c|c|c|}\hline
\ $Q^2$(GeV$^2$) \ & \ $\langle\hat{x}\rangle_q$ (n=1) \ & \ $\langle\hat{x}\rangle_q$ (n=2) \ & \ $\langle\hat{x}\rangle_q$ (n=3) \ & \ $\langle\hat{x}\rangle_q$ (n=4) \ & \ $\langle\hat{x}\rangle_q$ (n=5)\ \\ \hline
$Q^2=Q_0^2$ & $3.7\times10^{-2}$ & $-1.63\times10^{-1}$ & $5.07\times10^{-1}$ & -1.164 & 2.170 \\ 
10 & 2.781$\times10^{-1}$ & -9.52$\times10^{-1}$ & 2.527 & -4.950 & 8.107\\ 
40 & 3.582$\times10^{-1}$ & -1.112 & 2.830 & -5.329 & 8.5330 \\ 
60 & 3.750$\times10^{-1}$ & -1.150 & 2.897 & -5.399 & 8.6050 \\ 
80 & 3.911$\times10^{-1}$ & -1.176 & 2.939 & -5.455 & 8.660 \\ 
100 & 4.037$\times10^{-1}$ & -1.194 & 2.969 & -5.467 & 8.672 \\ \hline
\end{tabular}
\end{center}
\end{table}

In column 2 of Table \ref{t1}, we record the result of Ref\cite{DK5}, taking only one term of the infinite series. In the same table, we now show the corresponding results, taking upto 2, 3, 4, 5 terms of the two infinite series. From column 3 to 6, it is seen, partial momentum fractions carried by quarks are either -ve or exceed the theoretical limit.

\begin{figure}[!tbp]
\centering
\includegraphics*[width=90mm]{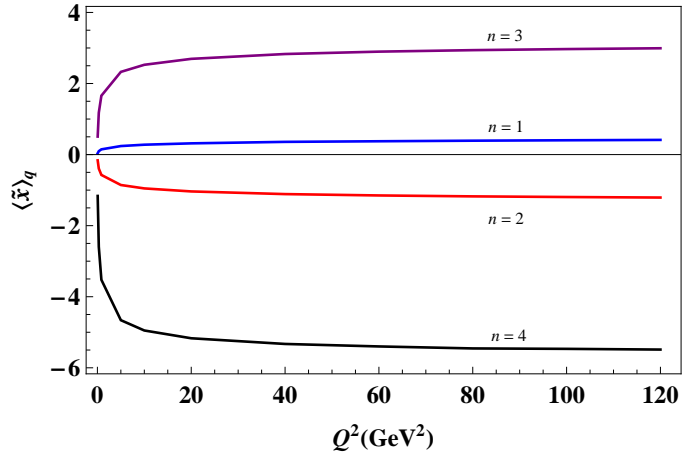}
\caption{For different values of \textit{n}, $\langle\hat{x}\rangle_q$ vs $Q^2$ (GeV$^2$) of \textbf{Model 1}}
\label{F5}
\end{figure}

In Fig \ref{F5}, we show the results of Table \ref{t1} graphically. It shows that the approximation used in Ref\cite{DK4} is not reasonable and an improved method is necessary.

\subsection{Semi-analytical and Numerical results of Model 1:}
\label{Bb}
As a consequences of the limitation of the analytical method, we take recourse to semi-analytical method i.e. we evaluate $I_1$ and $I_2$ numerically and then calculate $\langle\hat{x}\rangle_q$ by using Eqn(\ref{E18}), for a few representative values of $Q^2$(GeV$^2$). For the same values of $Q^2$, $\langle\hat{x}\rangle_q$ is calculated numerically by using Eqn(\ref{E15}). In Eqn(\ref{E15}), $e_i$ is the fractional electric charges of quarks and anti quarks. If we assume their flavored dependence and take number of flavors $n_f=4$, we obtain
\begin{equation}
\label{E25}
\sum_{i=1}^{4}e^{2}_{i} = \frac{10}{9}
\end{equation}
for u, d, s and c quarks leading to 
\begin{equation}
\label{E26}
\langle\hat{x}\rangle_q = \frac{9}{10} \int_{x_a}^{x_b} F_{2}(x,Q^2)dx
\end{equation}
\\
Similarly, for $n_f=5$ i.e. for  u, d, s, c and b quarks, we will have 

\begin{equation}
\label{E27}
\sum_{i=1}^{5}e^{2}_{i} = \frac{11}{9}
\end{equation}
and
\begin{equation}
\label{E28}
\langle\hat{x}\rangle_q = \frac{9}{11} \int_{x_a}^{x_b} F_{2}(x,Q^2)dx
\end{equation}
\\
\begin{table}[!bp]
\caption{}\label{t2}%
Results of $\langle\hat{x}\rangle_q$ of \textbf{Model 1} for semi-analytical, numerical method and upper limit of $\langle\hat{x}\rangle_g$ for numerical method for different $Q^2$
\begin{center}
\begin{tabular}{|c|c|c|c|c|c|}\hline
\ $Q^2$\ & \ $\langle\hat{x}\rangle_q$ \ & \ $\langle\hat{x}\rangle_q$ ($n_f=4$) \ & \ $\langle\hat{x}\rangle_g$ \ & \ $\langle\hat{x}\rangle_q$ ($n_f=5$) \ & \ $\langle\hat{x}\rangle_g$  \ \\ 
(GeV$^2$) & \ (semi-analytical) \ & (numerical) & \ ($n_f=4$) \  & (numerical) & \ ($n_f=5$) \ \\ \hline
\ $Q^2=Q_0^2$ \ & 1.941$\times10^{-4}$ & 6.063$\times10^{-4}$ & \ 9.993$\times10^{-1}$ \ & 5.576$\times10^{-4}$ & \ 9.994$\times10^{-1}$ \ \\
10 &  4.020$\times10^{-3}$ & 6.179$\times10^{-3}$ & 9.938$\times10^{-1}$ & 5.603$\times10^{-3}$ & \ 9.943$\times10^{-1}$ \ \\
40 &  7.549$\times10^{-3}$ & 8.857$\times10^{-3}$ & 9.911$\times10^{-1}$ & 8.058$\times10^{-3}$ & \ 9.919$\times10^{-1}$ \ \\
60 &  9.026$\times10^{-3}$ & 9.791$\times10^{-3}$ & 9.902$\times10^{-1}$ & 8.897$\times10^{-3}$ & \ 9.911$\times10^{-1}$ \ \\
80 &  1.023$\times10^{-2}$ & 1.050$\times10^{-2}$ & 9.895$\times10^{-1}$ & 9.548$\times10^{-3}$  & \ 9.904$\times10^{-1}$ \ \\
120 & 1.226$\times10^{-2}$ & 1.152$\times10^{-2}$ & 9.884$\times10^{-1}$ & 1.050$\times10^{-2}$ & 9.895$\times10^{-1}$ \\ \hline
\end{tabular}
\end{center}
\end{table}

In Table \ref{t2}, column 2 represents $\langle\hat{x}\rangle_q$ for semi-analytical method, while column 3 and 5 represents $\langle\hat{x}\rangle_q$ for numerical method in terms of $n_f=$ 4 and 5 respectively. Here, $\langle\hat{x}\rangle_q$ is recorded for $Q^2$ up to 120 GeV$^2$. It shows that the numerical values of improved results are not significantly different from those of Ref\cite{DK5}, presumably due to effective cancellation of odd and even terms of the infinite series. From Table \ref{t2}, we observe that as in Ref\cite{DK5}, in the improved analysis too, $\langle\hat{x}\rangle_q$ increases with the increasing $Q^2$. On the other hand, column 4 and 6 represents the upper limit of $\langle\hat{x}\rangle_g$ for $n_f=4$ and 5, calculated by using Eqn(\ref{E16}). It also decreases with the corresponding increasing $Q^2$ as in \cite{DK4}.

\subsection{Numerical results of Model 2:}
\begin{table}[!bp]
\caption{}\label{t3}%
Results of $\langle\hat{x}\rangle_q$ and upper limit of $\langle\hat{x}\rangle_g$ for $n_f=4$ and 5 of \textbf{Model 2} for different $Q^2$
\begin{center}
\begin{tabular}{|c|c|c|c|c|}\hline
\ $Q^2$\ & \ $\langle\hat{x}\rangle_q$ ($n_f=4$) \ & \ $\langle\hat{x}\rangle_g$ \ & \ $\langle\hat{x}\rangle_q$ ($n_f=5$) \ & \ $\langle\hat{x}\rangle_g$  \ \\ 
(GeV$^2$) \ & (numerical) & \ ($n_f=4$) \ & (numerical) \ & \ ($n_f=5$) \ \\ \hline
\ $Q^2=Q_0^2$ \ & 2.297$\times10^{-4}$ & 9.997$\times10^{-1}$ & 2.087$\times10^{-4}$ & 9.997$\times10^{-1}$ \\
2 &  3.539$\times10^{-3}$ & 9.964$\times10^{-1}$ & 3.217$\times10^{-3}$ & 9.967$\times10^{-1}$ \\
4 &  8.587$\times10^{-3}$ & 9.914$\times10^{-1}$ & 7.816$\times10^{-3}$ & 9.921$\times10^{-1}$ \\
6 &  1.455$\times10^{-2}$ & 9.854$\times10^{-1}$ & 1.328$\times10^{-2}$ & 9.867$\times10^{-1}$ \\
8 &  2.120$\times10^{-2}$ & 9.788$\times10^{-1}$ & 1.922$\times10^{-2}$ & 9.807$\times10^{-1}$ \\
10 &  2.833$\times10^{-2}$ & 9.716$\times10^{-1}$ & 2.566$\times10^{-2}$ & 9.743$\times10^{-1}$ \\ \hline
\end{tabular}
\end{center}
\end{table}

Here, we have calculated $\langle\hat{x}\rangle_q$ for model 2 numerically under the same procedure as done for model 1 in section \ref{Bb} by using the number of flavors $n_f= $ 4 and 5. The calculated results are given in Table \ref{t3} for $Q^2$ up to 10 GeV$^2$. Here too, we can see $\langle\hat{x}\rangle_q$ increases with increasing $Q^2$. In column 3 and 5, corresponding upper limit of $\langle\hat{x}\rangle_g$ for $n_f=4$ and 5 are given, which is calculated by using the Eqn(\ref{E16}) and it decreases as $Q^2$ increases.

\subsection{Numerical results of recent model of Block \textit{et.al.} (Model 3):}
For the comparison of improved results of the present model, we choose a more recent phenomenologically successful model suggested by Block, Durand, Ha and McKay \cite{blo}. The model has wide range of phenomenological validity in $Q^2$: $0.85\leq Q^2\leq 3000$GeV$^2$ for small $x\leq x_p=0.11$ which has Froissart Saturation like behavior \cite{fe}. To estimate the partial momentum fraction carried by quarks $\langle\hat{x}\rangle_q$ in the present range $6.2\times 10^{-7}\leq x\leq 10^{-2}$, we need to extract the quarks and anti quarks parton distribution function as defined
\begin{equation}
\label{E29}
F_2^p(x,Q^2)= x \sum_{i=1}^{n_f}e^{2}_{i} \left[  q_i(x,Q^2)+ \bar{q}_i(x,Q^2)\right] 
\end{equation}
\\
We can express $\langle\hat{x}\rangle_q$ by using Eqn(\ref{E15}): \textbf{(Model 3)}
\begin{equation}
\label{E30}
\langle\hat{x}\rangle_q = \left( \sum_{i=1}^{n_f}e^{2}_{i} \right)^{-1} \int_{x_a}^{x_b} F_{2}^p(x,Q^2)dx
\end{equation}

We will then see how $\langle\hat{x}\rangle_q$ changes with increasing $Q^2$. The expression for $F_2^p(x,Q^2)$ \cite{blo} is:
\begin{equation}
\label{E31}
F_2^p(x,Q^2) = (1-x)\left\lbrace \frac{F_p}{1-x_p}+A(Q^2)\ln\frac{x_p (1-x)}{x (1-x_p)}+ B(Q^2)\ln^2\frac{x_p (1-x)}{x (1-x_p)} \right\rbrace 
\end{equation}
Where,
\begin{eqnarray}
A(Q^2) = a_0 + a_1 \ln Q^2 + a_2 \ln^2 Q^2 \nonumber \\
B(Q^2) = b_0 + b_1 \ln Q^2 + b_2 \ln^2 Q^2
\end{eqnarray}
and the parameters fitted from deep inelastic scattering data \cite{blo} are
\begin{eqnarray}
x \leqslant x_p = 0.11 \ \ {\text a\text n\text d} \ \ F_p = 0.413 \pm 0.003 \ ,  \\ \nonumber
\end{eqnarray}
\begin{eqnarray}
\label{Ee}
a_0 &=& -8.471\times 10^{-2}\pm 2.62\times 10^{-3} \ , \nonumber \\
a_1 &=& 4.190\times 10^{-2}\pm 1.56\times 10^{-3} \ , \nonumber \\
a_2 &=& -3.976\times 10^{-3}\pm 2.13\times 10^{-4} \ , \nonumber \\
b_0 &=& 1.292\times 10^{-2}\pm 3.62\times 10^{-4} \ , \nonumber \\
b_1 &=& 2.473\times 10^{-4}\pm 2.46\times 10^{-4} \ , \nonumber \\
b_2 &=& 1.642\times 10^{-3}\pm 5.52\times 10^{-5} \ . \
\end{eqnarray}

In Table \ref{t4}, we record the numerical values of $\langle\hat{x}\rangle_q$ of Model 3 for $Q^2$ upto 3000 GeV$^2$ and also the upper limit of $\langle\hat{x}\rangle_g$ (using Eqn \ref{E16}) for the flavors  $n_f=$ 4 and 5 respectively.

\begin{figure}[!bp]
\centering
\includegraphics*[width=90mm]{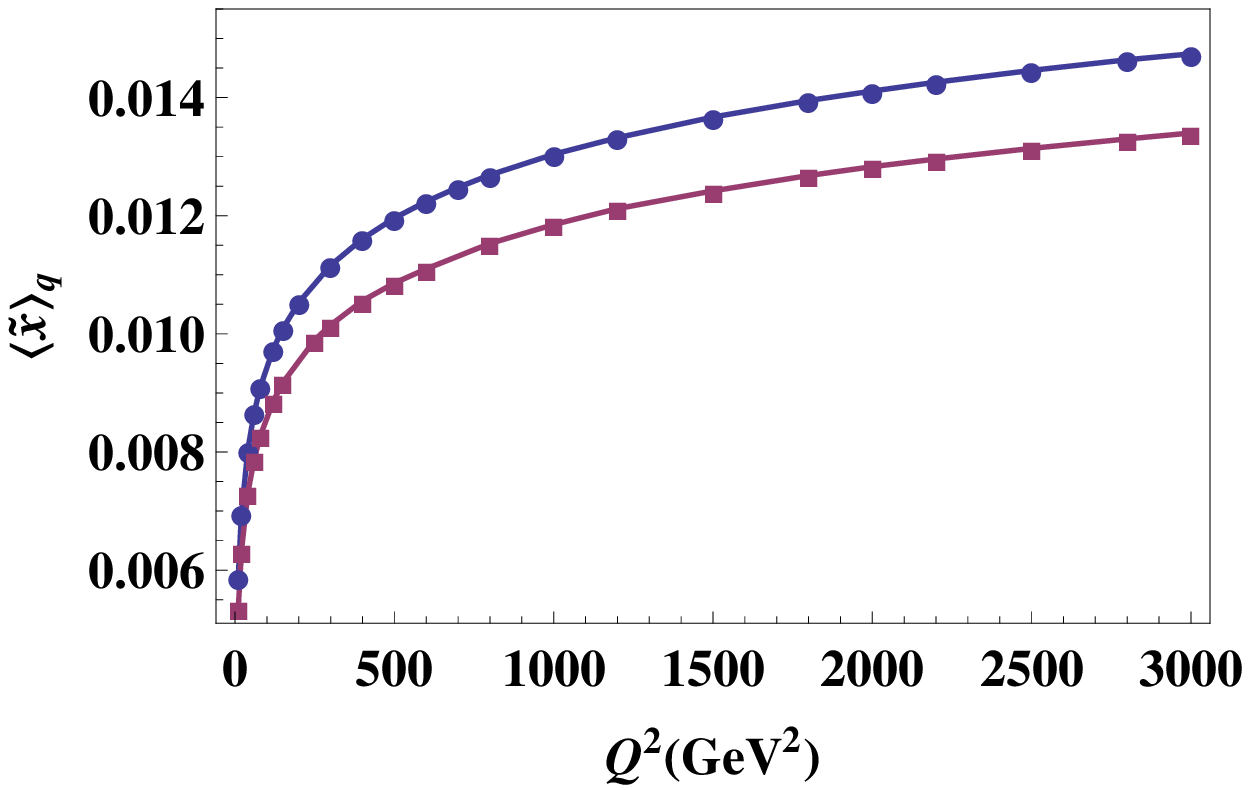}
\caption{$\langle\hat{x}\rangle_q$ vs $Q^2$ (GeV$^2$) of \textbf{Model 3} for $n_f=4$ (dots) and $n_f=5$ (squares)}
\label{F6}
\end{figure}

In Fig \ref{F6}, we have plotted the $\langle\hat{x}\rangle_q$  of model 3 for $n_f=4$ and 5. From Fig \ref{F6}, it can be seen, the rise of $\langle\hat{x}\rangle_q$ for $n_f=4$ is faster than that of $n_f=5$ i.e. the rise of $\langle\hat{x}\rangle_q$ becomes slower on increasing the number of flavors.

\begin{figure}[!tbp]
\centering
\includegraphics*[width=90mm]{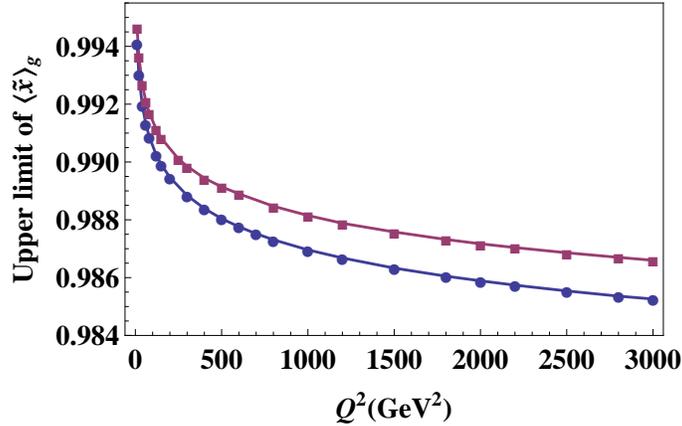}
\caption{Upper limit of $\langle\hat{x}\rangle_g$ vs $Q^2$ (GeV$^2$) of \textbf{Model 3} for $n_f=4$ (dots) and $n_f=5$ (squares)}
\label{F7}
\end{figure}

In Fig \ref{F7}, we have plotted the upper limit of $\langle\hat{x}\rangle_g$ of model 3 for $n_f=4$ and 5. Fig \ref{F7} shows that the upper limit of $\langle\hat{x}\rangle_g$ decreases as $Q^2$ increases and the fall is slower with the increasing $n_f$.

\begin{table}[!bp]
\caption{}\label{t4}%
Results of $\langle\hat{x}\rangle_q$ and upper limit of $\langle\hat{x}\rangle_g$ for $n_f=4$ and 5 of \textbf{Model 3} for different $Q^2$
\begin{center}
\begin{tabular}{|c|c|c|c|c|}\hline
\ $Q^2$ (GeV$^2$) \ & \ $\langle\hat{x}\rangle_q$ ($n_f=4$) \ & \ $\langle\hat{x}\rangle_g$ ($n_f=4$) \ & \ $\langle\hat{x}\rangle_q$ ($n_f=5$) \ & \ $\langle\hat{x}\rangle_g$ ($n_f=5$) \  \\ \hline
\ 0.85 \ & \ 2.051$\times10^{-3}$ \ & 9.979$\times10^{-1}$ & \ 1.865$\times10^{-3}$ \ & 9.981$\times10^{-1}$ \\
60 &  8.667$\times10^{-3}$ & 9.913$\times10^{-1}$ & 7.879$\times10^{-3}$ & 9.921$\times10^{-1}$ \\
150 & 1.009$\times10^{-2}$ & 9.899$\times10^{-1}$ & 9.174$\times10^{-3}$ & 9.908$\times10^{-1}$ \\
1500 & 1.367$\times10^{-2}$ & 9.863$\times10^{-1}$ & 1.242$\times10^{-2}$ & 9.875$\times10^{-1}$ \\
2000 & 1.411$\times10^{-2}$ & 9.859$\times10^{-1}$ & 1.283$\times10^{-2}$ & 9.871$\times10^{-1}$ \\
3000 & 1.474$\times10^{-2}$ & 9.852$\times10^{-1}$ & 1.340$\times10^{-2}$ & 9.866$\times10^{-1}$\\ \hline
\end{tabular}
\end{center}
\end{table}

\subsection{Comparison of pattern of $\langle\hat{x}\rangle_q$ for Model 1 and Model 3:}
\begin{figure}[!bp]
\centering
\includegraphics*[width=90mm]{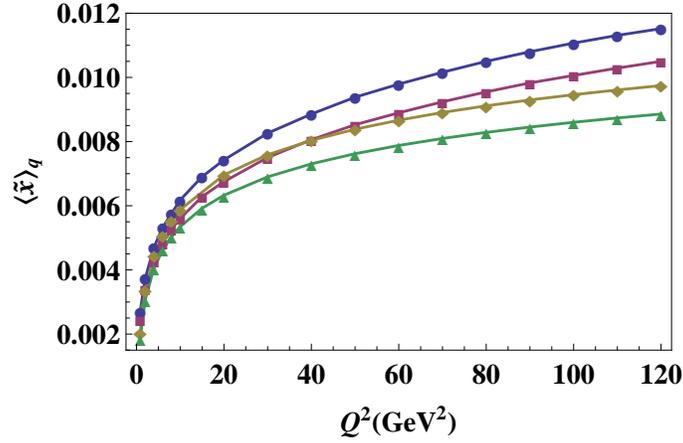}
\caption{$\langle\hat{x}\rangle_q$ vs $Q^2$ (GeV$^2$) of \textbf{Model 1} for $n_f=4$ (dots) and $n_f=5$ (squares) \\ $\langle\hat{x}\rangle_q$ vs $Q^2$ (GeV$^2$) of \textbf{Model 3} for $n_f=4$ (diamonds) and $n_f=5$ (triangles)
}
\label{F8}
\end{figure}

Here, we compare the pattern of $\langle\hat{x}\rangle_q$ of Models 1 and 3 for $n_f=4$ and 5 in Fig \ref{F8} by taking $Q^2$ upto 120 GeV$^2$. Model 2 is not taking into account, as it has a very restrictive range of $Q^2$ $\leqslant$ 10 GeV$^2$ only.

Fig \ref{F8} shows that for $n_f= 4$ and 5, the pattern of $\langle\hat{x}\rangle_q$ of both the Models 1 and 3 look similar. Also, the $\langle\hat{x}\rangle_q$ increases with the increasing $Q^2$, but the rise of Model 1 is faster than that of Model 3 for each flavors and the  $\langle\hat{x}\rangle_q$ for each models grows slower on increasing $n_f$ as expected.

\begin{figure}[!bp]
\centering
\includegraphics*[width=90mm]{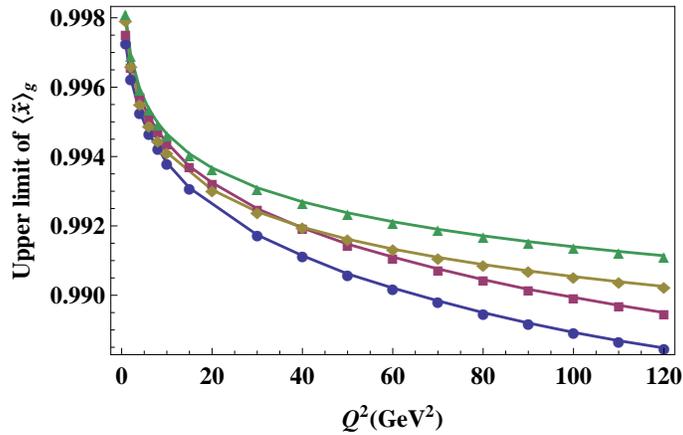}
\caption{The upper limit of $\langle\hat{x}\rangle_g$ vs $Q^2$ (GeV$^2$) of \textbf{Model 1} for $n_f=4$ (dots) and $n_f=5$ (squares) \\ $\langle\hat{x}\rangle_q$ vs $Q^2$ (GeV$^2$) of \textbf{Model 3} for $n_f=4$ (diamonds) and $n_f=5$ (triangles)
}
\label{F9}
\end{figure}

On the other hand, Fig \ref{F9} represents the  upper limit of $\langle\hat{x}\rangle_g$ vs $Q^2$ (GeV$^2$) of Models 1 and 3 for $n_f=$ 4 and 5. From Fig \ref{F9}, one can see the upper limit of $\langle\hat{x}\rangle_g$ falls down as $Q^2$ increases and the fall is slower with the increasing $n_f$ in each models. Also the fall of Model 1 is faster than that of Model 3.

\subsection{Comparison with perturbative and lattice QCD :}
Let us compare our results with the predictions of perturbative and lattice QCD.\\
In Ref \cite{dj, hd, al, rg}, the asymptotic QCD predictions of $\langle x \rangle_q$ and $\langle x
\rangle_g$ are:
\begin{equation}
\label{E32}
\lim_{Q^2\to\infty} \langle x \rangle_q = \frac{3n_f}{2n_g+3n_f},
\end{equation}
\begin{equation}
\label{E33}
\lim_{Q^2\to\infty} \langle x \rangle_g = \frac{2n_g}{2n_g+3n_f},
\end{equation}
\\
Here, $n_f$ and $n_g$ represent the number of active flavors and number of gluons respectively. For SU(3)$_c$, $n_g=8$. For $n_f$ = 5, Eqn(\ref{E32}) and (\ref{E33}) yield $\langle x \rangle_g$ = $\dfrac{1}{2}\left(\langle x\rangle_q +\langle x \rangle_g \right) $: 50 \% of the  momentum of  proton is carried by gluons, as noted in \cite{hd} and claimed  to  be experimentally tested in \cite{a}.

In Ref\cite{ch}, it has alternative asymptotic prediction:
 
\begin{equation}
\label{E34}
\lim_{Q^2\to\infty} \langle x \rangle_q = \frac{6n_f}{n_g+6n_f},
\end{equation}
\begin{equation}
\label{E35}
\lim_{Q^2\to\infty} \langle x \rangle_g = \frac{n_g}{n_g+6n_f},
\end{equation}
\\
Where Eqn(\ref{E32}) and (\ref{E33}) implies that except for $n_f = 6$, $\langle x\rangle_q < \langle x\rangle_g$. Specifically, for $n_f$ = 5, Eqn(\ref{E32}-\ref{E33}) yields $\langle x \rangle_g$ = $\dfrac{1}{2}$ $\left(\langle x\rangle_q +\langle x \rangle_g \right)$ and Eqn(\ref{E34}-\ref{E35}) gives $\langle x \rangle_g$ = $\dfrac{1}{5} \left(\langle x\rangle_q +\langle x \rangle_g \right)$. In the above equations, $\langle x \rangle_q$ and $\langle x \rangle_g$ denote the momentum fractions carried by quarks and gluons respectively for  the entire \textit{x}-range .

The difference between  Eqn(\ref{E32}-\ref{E33}) and Eqn(\ref{E34}-\ref{E35}) is attributed in Ref\cite{ch} to the proper  gauge invariant definition of gluon  momentum density; its definition  in earlier works \cite{dj, hd, al, rg} includes a quark - gluon interaction term  and  hence resulted in a inflated   value of gluon momentum fraction in proton.

However, later Ji\cite{ji} refutes the claim of Chen \textit{et al} \cite{ch}, underlying the correctness of the QCD prediction, Eqn(\ref{E32}-\ref{E33})  \cite{dj, hd, al, rg}.

However, none of the Ref\cite{dj, hd, al, rg, ch,ji} specifically states about the partial momentum fractions $\langle\hat{x}\rangle_q$ and $\langle\hat{x}\rangle_g$, relevant for phenomenological study in limited small \textit{x} regimes and finite $Q^2$, as in the present analysis. However, if the asymptotic predictions of Ref\cite{dj, hd, al, rg, ch,ji} are considered to be the rough indicator even for partial momentum fractions for small \textit{x} quarks at the finite $Q^2$, then the present result favors the prediction of Ref\cite{ch}, instead of Ref\cite{dj, hd, al, rg}, without violating the experimental value \cite{a}.

It is to be noted that the rise of partial momentum fraction of small \textit{x} quarks with $Q^2$ (specifically, the logarithmic rise with $Q^2$ in Model 3) is not necessarily inconsistent with the overall asymptotic prediction of total momentum fraction in perturbative QCD where $0<x<1$ \cite{dj, hd, al, rg}, if one analyzes the approximate $Q^2$ evolution of DGLAP equation at small \textit{x} \cite{c2} and large \textit{x} \cite{c1} respectively. At small \textit{x}, one finds that the quark evolves as ($\log Q^2)^n$ ($n$= {\scriptsize +}ve) while at large \textit{x}, the corresponding evolution will be ($\log Q^2)^{m}$ ($m<0$). Further, if one assumes
\begin{equation}
xG(x)= k(t)^\sigma F_2^s(x,Q^2)
\end{equation} 
where $t=\log\dfrac{Q^2}{\Lambda^2}$ \cite{b,c} and $k,\sigma \geq 0$, then both small \textit{x} quarks and gluon distribution will rise with $Q^2$ and the rise will be faster for the gluons than the quarks. However, in the context of the total momentum fraction of quarks and gluons, the small \textit{x} contribution is insignificantly smaller than the total and hence the intermediate and the large \textit{x }partons are the dominant contributor, resulting in the expected in perturbative QCD \cite{dj, hd, al, rg}. $\langle x\rangle_q$ falls with $Q^2$ rather than rises, as in the present analysis. It is also consistent with the well known result that the behavior of quarks and gluons at very small and large \textit{x} limit are \cite{b,c3} : \\
when $x\rightarrow 0$ , for small \textit{x} \cite{reg}
\begin{equation}
x f_i(x,Q^2 ) \longrightarrow x^{a_{f_i}(Q^2)} 
\end{equation}
for gluon
\begin{equation}
x f_g(x,Q^2 ) \longrightarrow x^{a_{f_g}(Q^2)} 
\end{equation}
and for large \textit{x}, when $x\rightarrow 1$ \cite{bro}
\begin{equation}
xf_i(x,Q^2 )  \longrightarrow (1 - x)^{b_{f_i}(Q^2)} 
\end{equation}
also for gluon
\begin{equation}
xf_g(x,Q^2 )  \longrightarrow (1 - x)^{b_{f_g}(Q^2)} 
\end{equation}
\\
Here $a_{f_g}$ is -ve and others are +ve.

At intermediate \textit{x} scale, one generally uses an interpolating function as polynomial \cite{c5} in \textit{x} $\sim \sum_{j=0}^{n} A_j x_j$. 

Taking into account all these aspects, it is therefore reasonable to expect that the total quark momentum fractions will fall with $Q^2$, while the corresponding gluon momentum fraction will rise leading to the expected QCD asymptotic prediction.\\ \\
Let us  discuss our results in the  context of lattice QCD \cite{md}.\\
Its  predictions for total momentum fractions for individual flavor are $\langle x\rangle_u$ = 34\%, $\langle x\rangle_d$ = 16\%, $\langle x\rangle_s$ = 4\% leading to total $\langle x\rangle_q$ = 54\%. The lattice analysis also yields $\langle x\rangle_g$ = 36\%, while remaining 10\% proton momentum fraction remained unaccounted. Thus, the analysis does not yet rule out the possibility of $\langle x\rangle_q$ that exceeds $\langle x\rangle_g$ at low momentum scale of lattice QCD, where perturbative QCD is not applicable. Such possibility is also not rule out in the present analysis at low $Q^2$.

\section{Conclusions}
\label{D}
In this paper, we have reported partial momentum fractions carried by small \textit{x} quarks in a limited \textit{x} ranges of $x_a<x<x_b$, first using phenomenological models of proton structure functions based on self-similarity and valid in limited ranges of small \textit{x}. Using momentum sum rule in inequality form, we obtain upper bound on momentum fractions carried by gluons in the same \textit{x}-ranges. We then compare the predictions of the self-similarity based models with the predictions of the QCD based phenomenologically successful model proposed by Block, Durand, Ha and McKay \cite{blo}

In each model, partial momentum fractions of small \textit{x} quarks increases with $Q^2$ in various degrees; liner rise in Models 1 and 2 and logarithmic rise in Model 3. However, upper bound of gluon partial momentum fractions invariably far exceed them. We then compared the predictions with perturbative, as well as lattice QCD expectation.

We have then suggested that such pattern of rise for small \textit{x} quarks (specifically, for the Model 3) is not inconsistent with known prediction of perturbative QCD. At lower $Q^2$ scale, Lattice QCD estimation of small \textit{x} quarks and gluons momentum fraction might also relevant. \\
\\
Let us conclude the paper with a few comments regarding the self-similarity as a relevant symmetry of structure of the proton, as used in Models 1 and 2.

The notion of self-similarity, although very interesting and has been successfully applied in hadron multi-particle production process, is not yet established in perturbative QCD: The experimental study during last decade has not yet confirmed this idea. Of course, some constraints from general approaches such as unitarity, analyticity and in particular the Froissart theorem \cite{fe} can be suitably incorporated in a self-similar proton, as been done in Ref \cite{dkc}. In a recent study \cite{bs1}, it is suggested that the logarithmic rise in $Q^2$ of structure function is also possible, in an improved singularity free self-similar based model with a wider kinematical range in \textit{x} and $Q^2$ but even then it is far short off accounting the entire $x-Q^2$ range explored in HERA unlike perturbative QCD \cite{HERA}. Further, such phenomenological models have got no predictive power outside their ranges of validity. Thus, it appears that the conjecture of self-similarity as a symmetry in the structure of the proton appears to have limited applicability, at least at the present experimental energy scale. 




\section*{Acknowledgment}
Final part of this work was completed when one of us (DKC) visited the Rudolf Peirels Center of Theoretical Physics, University of Oxford. He thanks Professor Subir Sarkar for hospitality. A other author (BS) acknowledges the UGC-RFSMS for financial support.


\begin{thebibliography}{99}

\bibitem{dj} David J Gross and Frank Wilczek, \textit{Phys.Rev. D} \textbf{9}, 980 (1974)
\bibitem{hd} H D Politzer, \textit{Phys.Rev.}\textit{D} \textbf{9}, 2174 (1974)
\bibitem{al} G Altarelli, \textit{Phys.Rep.} \textbf{81}, 1 (1982)
\bibitem{rg} R G Roberts, \textit{The Structure of the proton: Deep inelastic scattering}, Cambridge University Press, 1994
\bibitem{ch} Xiang-Song Chen \textit{et al.}, \textit{Phys. Rev. Lett.} \textbf{103}, 062001 (2009), hep-ph/0904.0321
\bibitem{ji} Xiangdong Ji, \textit{Phys. Rev. Lett.} \textbf{106}, 259101 (2011), hep-ph/0910.5022
\bibitem{md} M Deka \textit{et al.}, \textit{Phys. Rev. D } \textbf{91}, 1 (2015), hep-lat/1312.4816, 
\bibitem{b1} R S Bhalerao, \textit{Phys.Lett. B}\textbf{380}, 1 (1996), hep-ph/9607315
\bibitem{b2} H Navelet \textit{et.al.} \textit{Phys. Lett. B}\textbf{385}, 357 (1996)
\bibitem{b3} M Bertini Thesis,\textit{ Phenomenological study of Nucleon Structure Function,} (1995)\\ \url{http://www.iaea.org/inis/collection/NCLCollectionStore/_Public/27/049/27049057.pdf}
\bibitem{b4} A Watanabe and K Suzuki, \textit{Phys.Rev. D}\textbf{86}, 035011 (2012)
\bibitem{b5} A Watanabe and K Suzuki, \textit{Phys.Rev. D}\textbf{89}, 115015 (2014)
\bibitem{b6} A Watanabe and K Suzuki, \textit{``Proceedings of the 10th International Workshop on the Physics of Excited Nucleons (NSTAR2015)"}, DOI: 10.7566/JPSCP.10.052001, hep-ph/1510.00153
\bibitem{Last} T Lastovicka, \textit{Euro. Phys. J. C} \textbf{24}, 529 (2002), hep-ph/0203260
\bibitem{bb} B B Mandelbrot, \textit{Fractal Geometry of Nature}, W H Freeman, New York (1982)
\bibitem{kro} H Kr{\"o}ger, \textit{Phys. Reports} \textbf{323}, 81-181 (2000)
\bibitem{5} A. Bialas and R. Peschanski, \textit{Nucl. Phys. B} \textbf{703}, 273 (1986) \\ Ibid \textbf{857}, 308 (1988)
\bibitem{6} D. Ghosh \textit{et.al.}, \textit{Phys. Rev. D} \textbf{46} (1992)
\bibitem{7} Wu Yuanfang and Liu Lianshou, \textit{Int. J. Mod.Phys. A} \textbf{18}, 5337 (2003)
\bibitem{8} J D Bjorken, SLAC-PUB-6477 (1994)
\bibitem{DK4} A Jahan and D K Choudhury, \textit{Mod. Phys. Lett. A} \textbf{27}, 1250193 (2012), hep-ph/1304.6882
\bibitem{DK5} A Jahan and D K Choudhury, \textit{Mod. Phys. Lett. A} \textbf{28}, 1350056 (2013), hep-ph/1306.1891
\bibitem{bs} B Saikia and D K Choudhury, hep-ph/1409.0397 , \\ B Saikia and D K Choudhury, \textit{``Proceedings of National Conference on CICAHEP, Dibrugarh, 01, 149 (2015)"} \url{http://www.cicahep.org/?p=182}
\bibitem{fe} M Froissart, \textit{Phys. Rev.} \textbf{123}, 1053 (1961)
\bibitem{blo} M M Block, L Durand, P Ha and D W McKay, \textit{Phys. Rev. D} \textbf{84}, 094010 (2011), hep-ph/1108.1232
\bibitem{H1} H1:C Adloff \textit{et al.}, \textit{Euro. Phys. J. C} \textbf{21}, 33-61 (2001), hep-ex/0012053
\bibitem{ZE} ZEUS: J Breitweg \textit{et al.}, \textit{Phys. Lett. B} \textbf{487}, 53 (2000), hep-ex/0005018
\bibitem{HERA} H1 and ZEUS Collaborations, F.D. Aaron\textit{ et al}., \textit{JHEP} \textbf{01}, 109 (2010), hep-ex/0911.0884 
\bibitem{clo} F E Close,\textit{ An Introduction to Quarks and Partons} (Academic Press, 1979), p.233
\bibitem{a} T Sloan, G Smadja and R Voss, \textit{Phys. Rep.} \textbf{162}, 45 (1988)
\bibitem{c2} D K Choudhury and J K Sarma, \textit{Pramana} \textbf{39}, 273 (1992) \\ J K Sarma, D K Choudhury and G K Medhi, \textit{Phys. Lett. B} \textbf{403}, 139 (1997) \\ Saiful Islam and  D K Choudhury,\textit{ Euro. Phys.J. C} \textbf{72}, 2257 (2012)
\bibitem{c1} D K Choudhury and P K Dhar, \textit{Ind. J. Phys}. \textbf{81} (2), 259 (2007) \\ Ibid \textbf{83} (12), 1013 (2009) \\ D K Choudhury and P K Dhar, hep-ph/ 0910.2334

\bibitem{b} C Lopez and F J Yndurain, \textit{Nucl. Phys. B}\textbf{ 171}, 231 (1980)
\bibitem{c} D K Choudhury and Neelakshi N K Borah, hep-ph/1508.06041
\bibitem{c3} R D Ball, E R Nocera and J Rojo, hep-ph/1604.00024
\bibitem{reg} T. Regge, \textit{Nuovo Cim}. \textbf{14}, 951 (1959)
\bibitem{bro} S. J. Brodsky and G. R. Farrar, \textit{Phys. Rev. Lett.} \textbf{31}, 1153 (1973)
\bibitem{c5} ZEUS and H1 Collaborations, H Abramowicz \textit{et. al}., \textit{Eur. Phys. J. C} \textbf{75}, 580 (2015), hep-ex/1506.06042
\bibitem{dkc} A Jahan and D K Choudhury, \textit{Phys. Rev. D} \textbf{89}, 014014 (2014), hep-ph/1401.4327
\bibitem{bs1} D K Choudhury and B Saikia, hep-ph/1605.01149

\end{thebibliography}
\end{document}